\documentclass[preprint2]{aastex63}
\usepackage[utf8]{inputenc}
\usepackage{graphicx}
\usepackage{txfonts}
\usepackage[scaled]{helvet}
\usepackage{epsfig}
\usepackage{longtable}
\usepackage{url}
\usepackage[normalem]{ulem}
\usepackage{color}
\bibpunct{(}{)}{;}{a}{}{,}
\newcommand{\mj}{\ensuremath{\,M_{\rm J}}\space}

\newcommand{\ms}{\,m\,s$^{-1}$} 

\newcommand{\minus}{\scalebox{0.75}[1.0]{$-$}}

\usepackage{wrapfig}

\begin{document}

\title{Stellar Activity Manifesting at a One Year Alias Explains Barnard b as a False Positive}

\author[0000-0001-8342-7736]{Jack Lubin}
\affil{Department of Physics and Astronomy, University of California, Irvine, 4129 Frederick Reines Hall, Irvine, CA 92697, USA}

\author[0000-0003-0149-9678]{Paul Robertson}
\affil{Department of Physics and Astronomy, University of California, Irvine, 4129 Frederick Reines Hall, Irvine, CA 92697, USA}

\author[0000-0001-7409-5688]{Gudmundur Stefansson}
\altaffiliation{Henry Norris Russell Fellow}
\affil{Department of Astrophysical Sciences, Princeton University, 4 Ivy Lane, Princeton, NJ 08540, USA}

\author[0000-0001-8720-5612]{Joe Ninan}
\affil{Department of Astronomy and Astrophysics, The Pennsylvania State University, 525 Davey Laboratory, University Park, PA 16802, USA}
\affil{Center for Exoplanets and Habitable Worlds, The Pennsylvania State University, 525 Davey Laboratory, University Park, PA 16802, USA}

\author[0000-0001-9596-7983]{Suvrath Mahadevan}
\affil{Department of Astronomy and Astrophysics, The Pennsylvania State University, 525 Davey Laboratory, University Park, PA 16802, USA}
\affil{Center for Exoplanets and Habitable Worlds, The Pennsylvania State University, 525 Davey Laboratory, University Park, PA 16802, USA}

\author[0000-0002-7714-6310]{Michael Endl}
\affil{McDonald Observatory and Center for Planetary Systems Habitability, The University of Texas at Austin, Austin, TX 78730, USA}

\author[0000-0001-6545-639X]{Eric Ford}
\affil{Department of Astronomy and Astrophysics, The Pennsylvania State University, 525 Davey Laboratory, University Park, PA 16802, USA}
\affil{Center for Exoplanets and Habitable Worlds, The Pennsylvania State University, 525 Davey Laboratory, University Park, PA 16802, USA}
\affil{Institute for Computational \& Data Sciences, The Pennsylvania State University, University Park, PA, 16802, USA}

\author[0000-0001-6160-5888]{Jason T. Wright}
\affil{Department of Astronomy and Astrophysics, The Pennsylvania State University, 525 Davey Laboratory, University Park, PA 16802, USA}
\affil{Center for Exoplanets and Habitable Worlds, The Pennsylvania State University, 525 Davey Laboratory, University Park, PA 16802, USA}

\author[0000-0001-7708-2364]{Corey Beard}
\affil{Department of Physics and Astronomy, University of California, Irvine, 4129 Frederick Reines Hall, Irvine, CA 92697, USA}

\author[0000-0003-4384-7220]{Chad Bender}
\affil{Steward Observatory, University of Arizona, 933 N Cherry Ave., Tucson, AZ 85721, USA}

\author[0000-0001-9662-3496]{William~D.~Cochran}
\affil{McDonald Observatory and Center for Planetary Systems Habitability, The University of Texas at Austin, Austin TX 78712, USA}

\author[0000-0002-2144-0764]{Scott A. Diddams}
\affil{Time and Frequency Division, National Institute of Standards and Technology, 325 Broadway, Boulder, CO 80305, USA}
\affil{Department of Physics, University of Colorado, 2000 Colorado Avenue, Boulder, CO 80309, USA}

\author[0000-0002-0560-1433]{Connor Fredrick}
\affil{Associate of the National Institute of Standards and Technology, Boulder, Colorado 80305, USA}
\affil{Department of Physics, University of Colorado, Boulder, Colorado 80309, USA}

\author[0000-0003-1312-9391]{Samuel Halverson}
\affil{Jet Propulsion Laboratory, California Institute of Technology, 4800 Oak Grove Drive, Pasadena, CA 91109, USA}

\author[0000-0001-8401-4300]{Shubham Kanodia}
\affil{Department of Astronomy and Astrophysics, The Pennsylvania State University, 525 Davey Laboratory, University Park, PA 16802, USA}
\affil{Center for Exoplanets and Habitable Worlds, The Pennsylvania State University, 525 Davey Laboratory, University Park, PA 16802, USA}

\author[0000-0001-5000-1018]{Andrew J. Metcalf}
\affil{Space Vehicles Directorate, Air Force Research Laboratory, 3550 Aberdeen Ave SE, Kirtland AFB, NM 87117, USA}
\affil{Time and Frequency Division, National Institute of Standards and Technology, 325 Broadway, Boulder, CO 80305, USA}
\affil{Department of Physics, University of Colorado, 2000 Colorado Avenue, Boulder, CO 80309, USA}

\author[0000-0000-0000-0000]{Lawrence Ramsey}
\affil{Department of Astronomy and Astrophysics, The Pennsylvania State University, 525 Davey Laboratory, University Park, PA 16802, USA}
\affil{Center for Exoplanets and Habitable Worlds, The Pennsylvania State University, 525 Davey Laboratory, University Park, PA 16802, USA}

\author[0000-0001-8127-5775]{Arpita Roy}
\affil{Space Telescope Science Institute, 3700 San Martin Drive, Baltimore, MD 21218, USA}
\affil{Department of Physics and Astronomy, Johns Hopkins University, 3400 N Charles St, Baltimore, MD 21218, USA}

\author[0000-0002-4046-987X]{Christian Schwab}
\affil{Department of Physics and Astronomy, Macquarie University, Balaclava Road, North Ryde, NSW 2109, Australia}

\author[0000-0002-4788-8858]{Ryan Terrien}
\affil{Department of Physics and Astronomy, Carleton College, One North College Street, Northfield, MN 55057, USA}

\begin{abstract}
Barnard's star is among the most studied stars given its proximity to the Sun. It is often considered \textit{the} Radial Velocity (RV) standard for fully convective stars due to its RV stability and equatorial declination. Recently, an $M \sin i = 3.3  M_\earth$ super-Earth planet candidate with a 233 day orbital period was announced by \citet{Ribas2018}. New observations from the near-infrared Habitable-zone Planet Finder (HPF) Doppler spectrometer do not show this planetary signal. We ran a suite of experiments on both the original data and a combined original + HPF data set. These experiments include model comparisons, periodogram analyses, and sampling sensitivity, all of which show the signal at the proposed period of 233 days is transitory in nature. The power in the signal is largely contained within 211 RVs that were taken within a 1000 day span of observing. Our preferred model of the system is one which features stellar activity without a planet. We propose that the candidate planetary signal is an alias of the 145 day rotation period. This result highlights the challenge of analyzing long-term, quasi-periodic activity signals over multi-year and multi-instrument observing campaigns.

\end{abstract}

\keywords{Barnard's Star, GJ 699, Barnard b, Exoplanets, Stellar Activity}

\section{Introduction}
\label{intro}

\par In the 105 years since its discovery by E.E. Barnard as the star with the largest proper motion \citep{BarnardDiscovery}, Barnard's star (GJ 699) has become one of the most studied and heavily scrutinized star systems. Adding to its distinguishing characteristics, it is the nearest star to our own Sun in the Northern Celestial Hemisphere, and the second-closest star system overall \citep{gaiaDR2,Bailerjones2018}. This proximity has engendered fascination by astronomers in many sub-fields of astronomy.

\par All these qualities have made the star attractive to astronomers in exoplanet science for decades. In 1963, Peter van de Kamp believed he had detected an astrometic wobble of Barnard's star using Swarthmore College's 24 inch refractor at Sproul Observatory \citep{vdkamp1}, which he attributed to a planet. He later updated his findings three more times, proposing a second planet in the system \citep{vdkamp2} and then revising the orbital parameters of both planets twice, finally concluding the system was comprised of 0.7 \mj and 0.5 \mj planets orbiting with periods of 12 and 20 years, respectively \citep{vdkamp3,vdkamp4}.

\par The 1975 and 1982 revisions were published in spite of two earlier results challenging the validity of those planets' detections. In the first, \citet{Hershey} used van de Kamp's photographic plates to determine that all the stars in the field of Barnard's star appeared to wobble in concert, and the source of this variability could be traced to telescope and instrumentation upgrades at times concurrent with the shifts in the data. Additionally, \citet{GateEich} reported no astrometric wobble of Barnard's star using the Van Vleek and Allegheny Observatories at Wesleyan University and University of Pittsburgh, respectively. More recent studies have confirmed that van de Kamp's planets are not recoverable by instruments which ought to have detected them easily \citep{Benedict1999, Kurster2003, Choi2013, Ribas2018}.

\par The Doppler, or radial velocity (RV), method for planet detection has its own challenges for low amplitude signals. Currently, the most prominent source of false positives is stellar activity. This activity, which results from phenomena such as starspots and plages, is modulated by the stellar rotation; it can induce RV signals that mask or masquerade as planets \citep{Robertson2014, Robertson2015}.  As our instruments become more precise, we are finding that even the quietest stars are indeed variable below 1 m\,s$^{-1}$. Furthermore, the interaction between stellar activity signals and other signals, such as planets, can worsen aliasing, which can occur when reconstructing a signal from incomplete sampling \citep{Robertson581}. The quasi-periodic nature of these stellar activity-induced signals, coupled with the inherently uneven sampling in astronomical observations, creates conditions for signal aliasing. Stellar activity is difficult to model and predict, but it has been shown that observing in the near-infrared (NIR) can help mitigate the effect of starspot dominated activity \citep{Marchwinski2015}; however, the amount of mitigation is limited based on target star's effective temperature and the impact of magnetic fields on line profiles changes \citep{Reiners2010}. Despite the problems that stellar activity induced signals create, these signals often decay away with time. Therefore, we are more confident in the planetary origin of a signal when it persists for much longer than the typical spot lifetime, although we have seen spot signals on M Dwarfs persist for longer than we might expect for Sunlike stars \citep{Robertson2020}.

\par When surveying old M dwarf stars like Barnard's star, the problem of stellar activity induced signals is exacerbated due to the lifetimes of spots. Starspots on G-type stars will typically survive $\sim$3 stellar rotations (depending on spot size) \citep{Giles2017}, but a starspot on the surface of an old M-Dwarf star might live through many more rotations of the star. \citep{RobertsonKapteyn, Robertson2020, Davenport2020}. Furthermore, while the average G-type star has a stellar rotation period of 10-20 days \citep{Nielson2013, McQuillan2014}, Old M dwarfs can have rotation periods in excess of 100 days \citep{Newton2016, SuarezMasc2018}. Therefore, we can expect starspot and active region RV signals on old M dwarf stars to maintain high signal power for hundreds to thousands of days. This creates a unique problem: such a time scale is long enough for signals to be measured across multiple consecutive observing seasons. When the stellar rotation period begins to rival a significant fraction of the observing season, the conditions for significant aliasing are firmly in place. Further complicating the matter for M Dwarfs, the periods at which stellar activity-induced signals exist---namely, the rotation period, its harmonics, and aliases---can coincide with the periods we would expect for a planet orbiting in or near the star's habitable zone \citep{Kopparapu2013, Vanderburg2016, Newton2018}.

\par Barnard's star is often referred to as a Doppler standard star for its relatively quiet nature, RV stability, high apparent brightness, and equatorial declination. Through this combination of factors, it is widely considered as \textit{the} RV standard for fully convective stars \citep[e.g.][]{Bean2010,anglada2012}. Such standard stars are vital to the exoplanet community for instrument commissioning and calibration. As our instruments become more precise, we may find planets and/or activity signals for \textit{all} stars. It is crucial that we fully understand and characterize the signals associated with whichever stars we designate as standards so that we can continue to have standard stars at all.

\par The Habitable-zone Planet Finder (HPF) began observing Barnard's star in the Spring of 2018 for engineering purposes. At the time, the star was not known to host a planet. Shortly after, \citet{Ribas2018} (hereafter R18) announced a super-Earth planet candidate orbiting Barnard's star with a 232.8 $\pm$ 0.4 day orbital period. \citet{Metcalf2019} presented early observations of the star to demonstrate the near-infrared RV precision achievable with HPF and its laser comb, but did not discuss the planet candidate due to the relatively short observational span compared to the period of the signal. As we continued to observe the star, we still did not find clear evidence of a signal at 233 days. This, combined with the findings from \citet{Kurster2003} which found a correlation between the RVs and H$\alpha$ values for Barnard's star, prompted us to revisit the full RV data set and the corresponding activity tracer time series.

\par After our own analysis of the original discovery data set and the addition of new data from HPF, we find that the signal at 233 days is transitory in nature and is an alias of the 145 $\pm$ 15 day rotation period as reported by \citet{ToledoP2019}: $\big(\frac{1}{145}-\frac{1}{365}\big)^{-1} = 240$. The 233 day proposed period is well within the 1$\sigma$ uncertainty on the rotation period. We are therefore compelled to conclude that the RV data of Barnard's star is best explained without the planet Barnard b proposed by R18.

\par This paper is organized as follows. In \S \ref{data}, we outline the data sources, split into the discovery data set (\S\ref{discdata}) and the subsequent modifications we made to create an updated data set (\S \ref{updateddata}). In \S\ref{analysis}, we perform the experiments of model comparison (\S\ref{modelcomparision}), periodogram analysis (\S\ref{periodograms}), and sampling sensitivity (\S\ref{samplesense}). In \S\ref{discussion}, we discuss the ramifications of this false positive result and in \S\ref{conclude} we conclude.

\section{Data}
\label{data}

\subsection{Discovery Data}
\label{discdata}

\par The claim by R18 of a 233-day planet candidate orbiting Barnard's star was based on a data set that was assembled over a 20 year span using multiple Doppler spectrometers. These instruments included those which employed the iodine-cell (I$_2$) method \citep{Valenti1995, Butler1996}: 186 points from the High Resolution \'Echelle Spectrometer \citep[HIRES;][]{Vogt1994} installed on the 10\,m Keck I Telescope at Maunakea in Hawaii; 43 points from the Automated Planet Finder \citep[APF;][]{Vogt2014} installed on the 2.4\,m telescope located at Lick Observatory on Mt. Hamilton outside of San Jose, California;  75 points from the UVES spectrograph installed on the 8.2\,m VLT UT2 at Paranal Observatory in Chile \citep{Dekker2000}; and 39 points from the Planet Finder Spectrometer \citep[PFS;][]{Crane2010} installed on the 6.5m Magellan II located at Las Campanas Observatory in La Silla, Chile. Other instruments used for data collection include 187 points and 40 points from the HARPS \citep{Mayor2003} and HARPS-N \citep{Cosentino2012} spectrometers respectively, installed on the ESO 3.6\,m at La Silla in Chile and the 3.5\,m Telescopio Nazionale Galelio at La Palma, and lastly 201 points from the visible channel of the CARMENES spectrograph \citep{Quirrenbach2016} installed on the 3.5\,m telescope at Calar Alto Observatory in Spain.

\par In all, 771 RV points were used. Further detailed information on the data sources and reduction methods can be found in R18. For ease of reference herein, we have dubbed this the \textit{discovery} data set.

\par In addition to RV data, R18 provided time series of Calcium II H\&K ($S_{HK}$) and H$\alpha$ indices for the instruments where either or both of these activity tracers are available. $S_{HK}$ and the H$\alpha$ index are known tracers of stellar activity which measure the filling in of the photospheric lines from heating in the chromosphere due to increased magnetic flux.

\subsection{Updates to Data}
\label{updateddata}

\par We made a few updates to the discovery data set. The first concerned the UVES H$\alpha$ time series. R18 published 21 H$\alpha$ points alongside 75 RV points, originally from \citet{Zechmeister2009b}. We noticed a non-uniform offset in time between H$\alpha$ epochs and seemingly corresponding RV epochs of $\sim$0.5 days. Therefore, we manually changed the H$\alpha$ timestamps to equal the nearest corresponding RV timestamp for consistency.

\par Next, we substituted more recent reductions of the RV data for certain instruments for the (now) older reductions used by R18. First, we used the velocities from the reduction of HIRES data provided by \citet{TalOr2019}. Additionally, we split the HIRES data into ``pre" and ``post" with respect to the instrument's CCD upgrade in August of 2004. This was motivated by \citet{TalOr2019}, who fit an offset, but we kept this as a free parameter when modelling for completeness. As several nights in this reduction contain multiple observations, we performed a nightly binning by taking a weighted average with weights $\sigma_{RV}^{-2}$. This binning preserved the sampling used by R18, except the updated time series includes one additional epoch at BJD = 2453301.74204. We excluded the last observation at 2456908.73079 since the associated velocity of 53.319 m/s makes it a $\sim$5.5 $\sigma$ outlier. We confirmed that this data set had been corrected for secular acceleration.

\par We similarly used the velocities from the reduction of HARPS data as performed by \citet{Trifonov2020}. We used the RVs which had been corrected for secular acceleration, nightly zero point offsets, and drift. Again, we maintained ``pre" and ``post" designations with respect to the HARPS fiber upgrade in 2015 for completeness \citep{LoCurto2015}. These data included more observations than were originally included in R18, with 108 observations occurring within an 11 day period in May 2013. These observations were taken as a part of the Cool Tiny Beats (CTB) program \citep{Berdinas2015}. Comparison between the CTB observations and the original HARPS data suggested that R18 performed a nightly averaging to compile their data set, but they do not describe exactly how they performed this binning. We chose to use a nightly weighted average with the same weights as described above.

\par
\begin{figure*}[t]
  \centering\includegraphics[width=\textwidth]{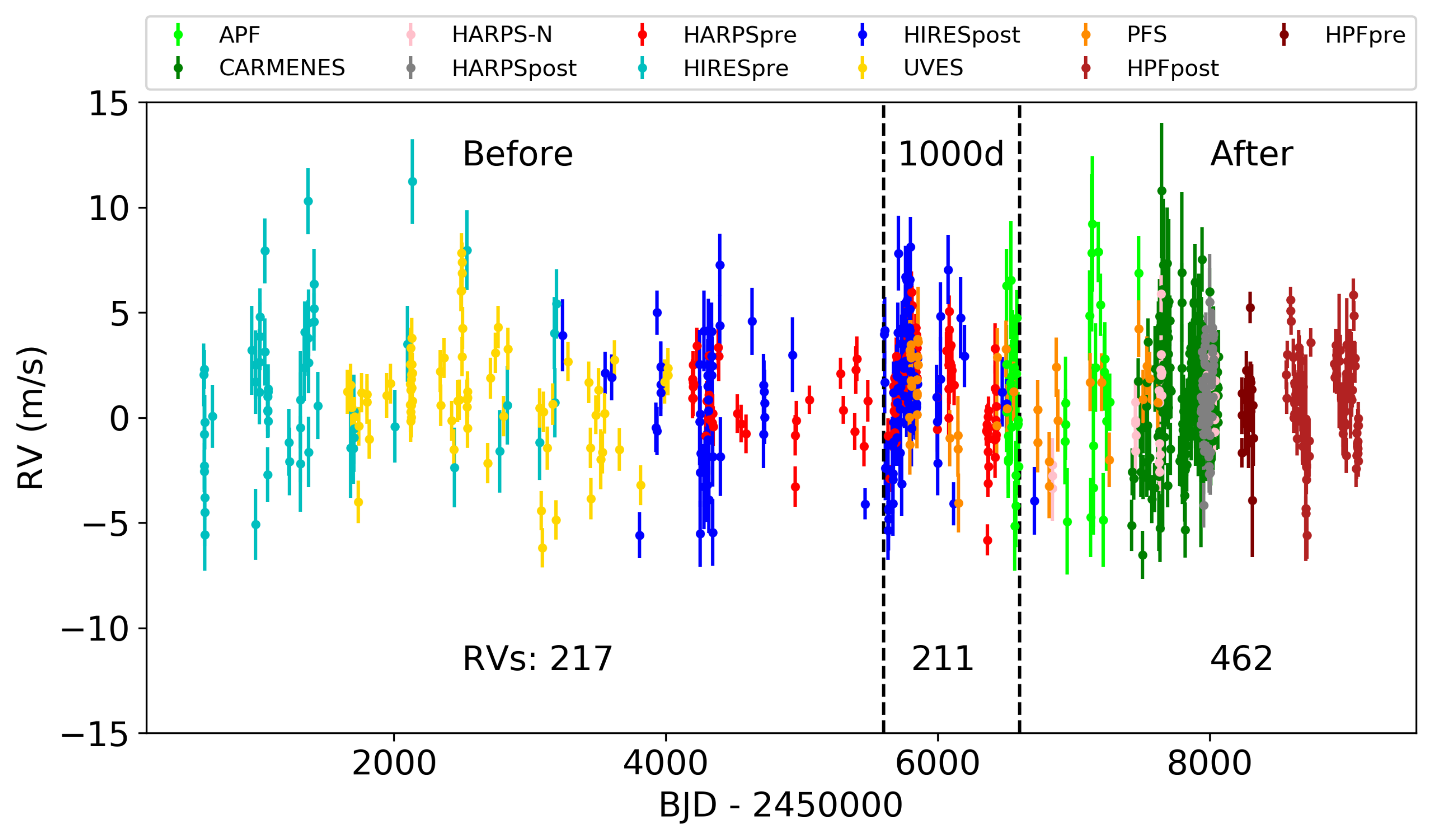}
 \centering\caption{The entire Updated RV time series spanning 23 years. Our three time windows of interest are noted along with their number of observations. Excluding HPF, the After window has 343 observations. Instrumental zero point offsets have been applied.}
  \label{fig:timeseries}
\end{figure*}

 The last modification to the discovery data set was the addition of the HPF data (\S\ref{sec:hpf_data}). For ease of reference herein, we have dubbed a new data set as the \textit{updated} data set. This updated data set includes the original, unaltered APF, CARMENES, HARPS-N, and PFS data, the UVES data with timestamps adjusted, the new data reductions of HIRES and HARPS (both split into pre and post domains), and the new HPF data, see Figure \ref{fig:timeseries}.

\begin{figure*}[t]
  \centering\includegraphics[width=\textwidth]{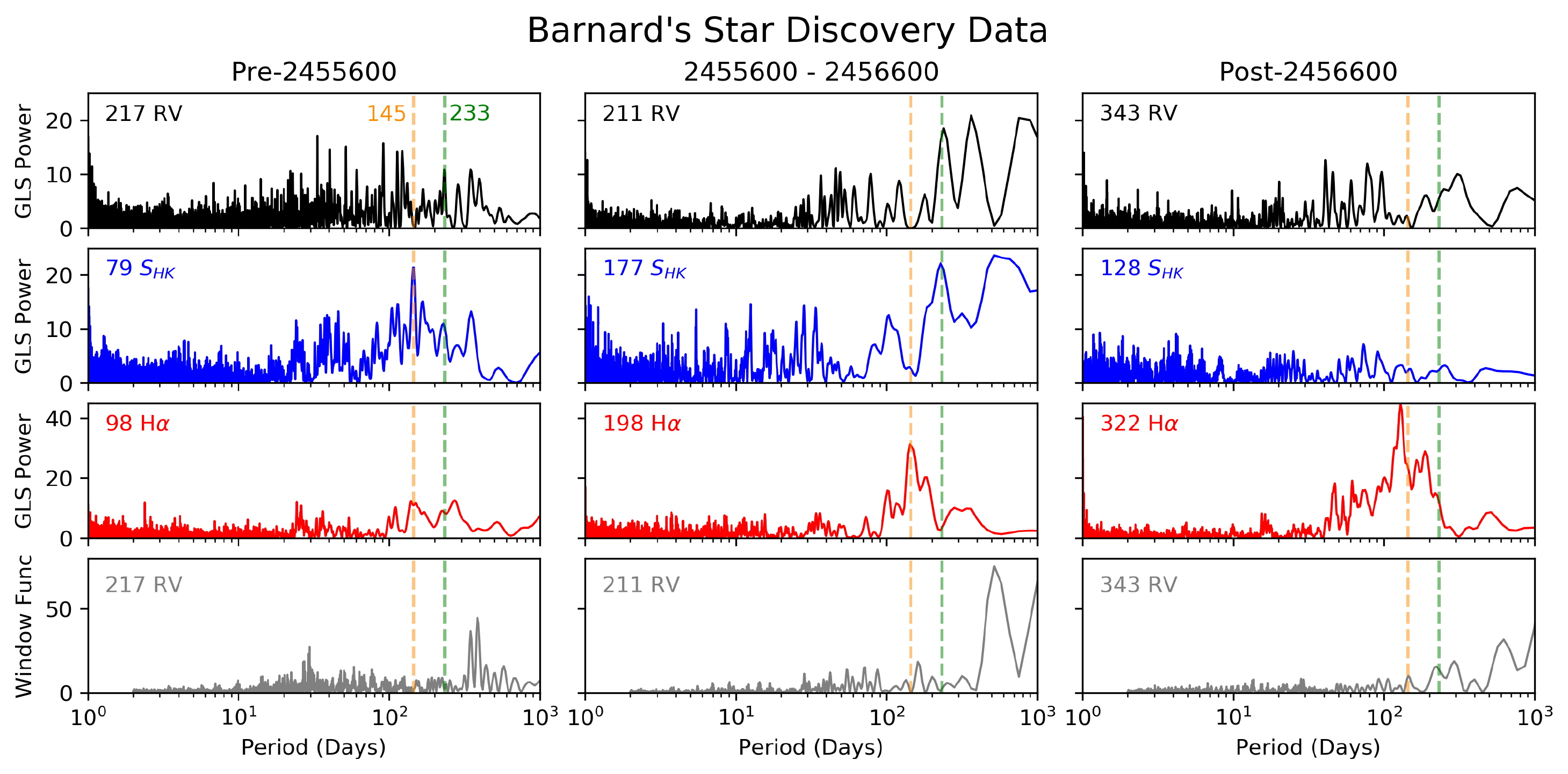}
  \centering\caption{GLS periodograms of the discovery data time series broken down in to three time windows. The middle column is the 1000d window. The 233 day signal is only present at significant power in the 1000d window in both the RVs (black) and CaHK S-values (blue). The H$\alpha$ Index (red) traces the rotation period. The Window Function for the RV time series in each window is the last row in gray. The orange line indicates 145 days, the stellar rotation period, and the green line indicates 233 days, the proposed planet period.}
  \label{fig:3x3}
\end{figure*}

\subsection{HPF Data}
\label{sec:hpf_data}

\par HPF is a high-resolution ($R\sim55,000$) near-infrared (NIR) spectrograph on the 10m Hobby-Eberly Telescope (HET), covering the Doppler information rich $z$, $Y$ and $J$ bands from 810---1280 nm \citep{Mahadevan2012, Mahadevan2014}. To enable precise RVs in the NIR, HPF is temperature stabilized to the sub-milliKelvin level \citep{Stefannson2016}. HET is a fully queue-scheduled telescope \citep{Shetrone2007}, and all observations were obtained as part of the HET queue. HPF has a NIR laser-frequency comb (LFC) calibrator which has been shown to enable $\sim$ 20 cm\,s$^{-1}$ calibration precision in 10 minute bins and 1.53 m\,s$^{-1}$ RV precision on-sky on Barnard’s star \citep{Metcalf2019} over an 3 month baseline. In this paper, we extend this baseline to 856 days. \cite{Stefansson2020} further discusses our drift correction algorithms. To enable maximum RV precision with HPF, we obtained all of our Barnard's star observations using the HPF LFC simultaneously with the on-sky observations.

\par To test the on-sky RV measurement performance of HPF, we observed Barnard's star as part of HPF Commissioning and ongoing Engineering time due to its brightness, overall known Doppler RV stability, and its rich RV information content in the NIR. Due to the restricted altitude design of the HET, the HET can only observe Barnard's star at certain times of the nights, or `tracks,' for approximately 68 minutes at a time. In total, we obtained 1016 high-quality spectra in 118 HET tracks with a median S/N of 479 per extracted 1D pixel evaluated at 1 $\mu$m. Due to a planned instrument thermal cycle in August 2018 which led to a minor RV offset before and after the cycling, we place an explicit RV offset before and after this event in our RV modeling. Before the thermal cycle, we generally obtained 6 exposures of Barnard's star with an exposure time of 300s in each HET track, while we generally obtained 10 exposures with an exposure time of 183s in each HET track after the thermal cycle. This change was made in order to harmonize the observing strategy of Barnard's star with that employed for the rest of the HPF 5-year blind Doppler survey. This standardized the observing for different targets according to their brightness, minimizing any potential risk of saturation for bright targets such as Barnard's star. The observing setup for Barnard's star has remained the same since the thermal break installation.

Overall, the 1016 high-quality spectra had a median exposure time of 191s, and a median photon-limited RV precision of 2.36\ms. After performing a weighted average of RV points within a given HET track, we obtain a median RV precision of 0.77\ms~per binned RV point. We use the binned RV points for subsequent analysis, which can be found in Table \ref{tbl:HPFRVs} in the appendix.

\par The 1D HPF spectra were reduced from the H2RG up-the-ramp data using the algorithms and procedures described in \cite{ninan2018}, \cite{kaplan2018}, and \cite{Metcalf2019}. Following the 1D spectral reduction, we calculated precise radial velocities of Barnard's star using an adapted version of \texttt{SERVAL} (SpEctrum Radial Velocity AnaLyzer) \cite{zechmeister2018}, optimized to analyze HPF spectra as further discussed in \cite{Metcalf2019} and \cite{Stefansson2020}. To derive precise RVs, \texttt{SERVAL} uses the template-matching algorithm, which has been shown to be particularly effective at producing precise radial velocities for M-dwarfs \citep{anglada2012}. To estimate accurate flux-weighted barycentric velocities, we used the \texttt{barycorrpy} package \citep{kanodia2018}, which uses the methodology and barycentric-correction algorithms presented in \cite{wright2014}. Following \cite{Metcalf2019} and \cite{Stefansson2020}, to derive the RVs we only use the 8 HPF orders cleanest of tellurics, covering the wavelength regions from 8540-8890\AA, and 9940-10760\AA. To minimize the impact of telluric and sky-emission lines on the RV determination, we explicitly mask out such lines as described in \cite{Metcalf2019} and \cite{Stefansson2020}. We subtracted the estimate sky background from the target spectra using the dedicated HPF sky fiber.

\section{Analysis}
\label{analysis}

\par We have taken a three-pronged approach to our investigation of the 233 day signal, largely performing the same experiments on both the discovery and updated data sets. First, we model the system's activity and/or potential planet with Gaussian Process (GP) regression models to identify a preferred model. Next, we perform a periodogram analysis to show the transitory nature of the signal. Finally, we analyze subsets of the data to show how the signal power is concentrated in time.

\subsection{Model Comparison}
\label{modelcomparision}

\par We began our analysis by modeling the stellar activity and/or planet signals for both the discovery and updated data sets. We chose to model stellar activity signals with GP regression because the GP framework has proven especially adept at modeling quasi-periodic signals associated with stellar rotation \citep{Haywood2014, Rajpaul2015, SuarezMasc2020, Bortle2021}.

\par In order to better constrain the stellar rotation signal, we first modeled the activity tracers using GP regression. We used a modified version of \texttt{Radvel} \citep{Fulton2018} which uses logarithmic priors for the GP hyperparameters, setting broad priors as advised by \citet{Angus2018}. We used the \texttt{Celerite} package \citep{celerite} for its efficiency.  Specifically, we use the celerite-compatible replacement for the quasi-periodic kernel for the covariance between the i$^{th}$ and j$^{th}$ observations:  \linebreak

\begin{center}
  $k_{ij} =$ \\
  $\frac{B}{2+C}\exp \left ( \frac{- \left | t_i -t_j\right |}{L} \right )\left [ \cos(\frac{2\pi \left | t_i -t_j\right |}{P_{rot}}) + (1+C)\right ] + \sigma ^{2}\delta _{ij},
  $\label{eq:kernel} \linebreak
\end{center}

equivalent to Equation 56 in \citet{celerite}. The hyperparameters of this kernel have similar interpretations to the quasi-periodic kernel function: $B$ is related to the signal amplitude, $L$ is the decay timescale for the exponential term in days, $C$ is a scaling term, $\sigma^{2}$ is the jitter from additional white noise beyond what can be accounted for in the formal measured uncertainties, and $P_{rot}$ is the recurrence timescale of the signal in days; in this case our astrophysical interpretation is closely related to the stellar rotation period. Each instrument received its own $B$ and $\sigma$ hyperparameters, but all instruments shared global $C$, $L$, and $P_{rot}$ hyperparameters.

\par We first performed a GP regression model to the discovery data set's H$\alpha$ values. Table \ref{tbl:halpha_model} shows our priors and subsequent maximal posteriors from this fit. The resulting posterior value for $\log C = -5.85$ indicates that the signal in question, i.e. the stellar rotation, is highly periodic in nature. \citet{ToledoP2019} states a rotation period of 145 $\pm$ 15 days, and from the results of our GP model of the H$\alpha$ data we are able to achieve a consistent result at $143.7^{+11.83}_{-10.93}$ days. Performing a GP with the same priors on hyperparameters for the $S_{HK}$ time series yielded similar results.

\par Using the H$\alpha$ posteriors of the $\log L$, $\log C$, and $P_{rot}$ terms as the priors in the RV model, we ran a suite of models on the RVs from the discovery data. In all of these RV models, the $\log C$ posterior value continued to prefer negligibly small values. We quickly found a similar behavior with the $\log B$ hyperparameter for APF as we had for the $\log C$ parameter. We believe that, due to the small aperture and iodine calibration on such a faint and red target star, the variability in the APF data is due primarily to photon noise rather than correlated astrophysical variability. We therefore opted to perform all analysis without a GP term for the APF RVs. When modelling activity, APF was allowed only an instrumental jitter and offset term, while when considering a planet, APF data were modelled with a Keplerian and instrumental jitter and offset only.

\par We computed three models on the discovery data set's RVs: a GP fit only, a 233d planet fit only, and a GP + 233d planet fit. Table \ref{tbl:disc_compare} in the Appendix shows the resulting Bayesian information criteria (BIC) and the number of free parameters in each model. We compare models by their $\Delta$BIC, preferring more complex models to simpler models when $BIC_{simple} - BIC_{complex} > 10$ \citep{kass1995}. For the discovery data, this criteria is satisfied for the GP-only model compared to the Planet-only model, but it is not satisfied for the GP+Planet compared to the GP-only. Thus the GP-only model is our preferred model. We then repeated this experiment of model comparisons using the updated data set. We used the same priors as for the discovery data set and we achieved a similar result, as shown in Table \ref{tbl:update_compare}. The GP+233d model has the smallest BIC, but the $\Delta$BIC between this model and GP only model does not justify the more complex model's additional free parameters, and so therefore the GP only model is once again our preferred model. The posteriors for all models using the discovery data can be found in Tables \ref{tbl:halpha_model} and \ref{tbl:disc_rv_models}, and using the updated data in Table \ref{tbl:updated_rv_models} in the Appendix.

\subsection{Periodograms}
\label{periodograms}

\par After finding the preferred model to be one which accounts for stellar activity only, we set out to determine how the activity signal might have revealed itself as a planetary signal in the GLS periodogram of the RV time series.

\par In an attempt to show coherence of the planetary signal, we split the discovery data into roughly equal thirds by observation number (taking care to not cut in the middle of an observing season) and computed the generalized Lomb-Scargle periodogram \citep[GLS;][]{Zechmeister2009} of each block of data. The 233 day signal is strongly present in the middle third, as shown in Figure \ref{fig:3x3}. Given some idealized data set which has comparable quality data, sampling, and white noise in all subsets of time, a true planetary signal must persist in all subsets. It will stay in phase across the entire observational time baseline, and the loss of power from removing a fixed number of data should be consistent regardless of which specific data are removed. While this data set is not such a perfect data set, the result of this test was the first indication that the 233d signal is not planetary in nature.

\par We then divided the corresponding $S_{HK}$ values in the same manner and, after computing GLS periodograms for each block, we saw similar behavior. Notably, in the time window when the RV signal at 233d is strongest, there is a matching peak in the $S_{HK}$ periodogram. This result is similar to \citet{Hatzes2013} where it was shown that activity signals changing over seasonal time spans can imprint planet-like periodicity into the RV time series.

\begin{figure}[t]
  \centering\includegraphics[width=.5\textwidth]{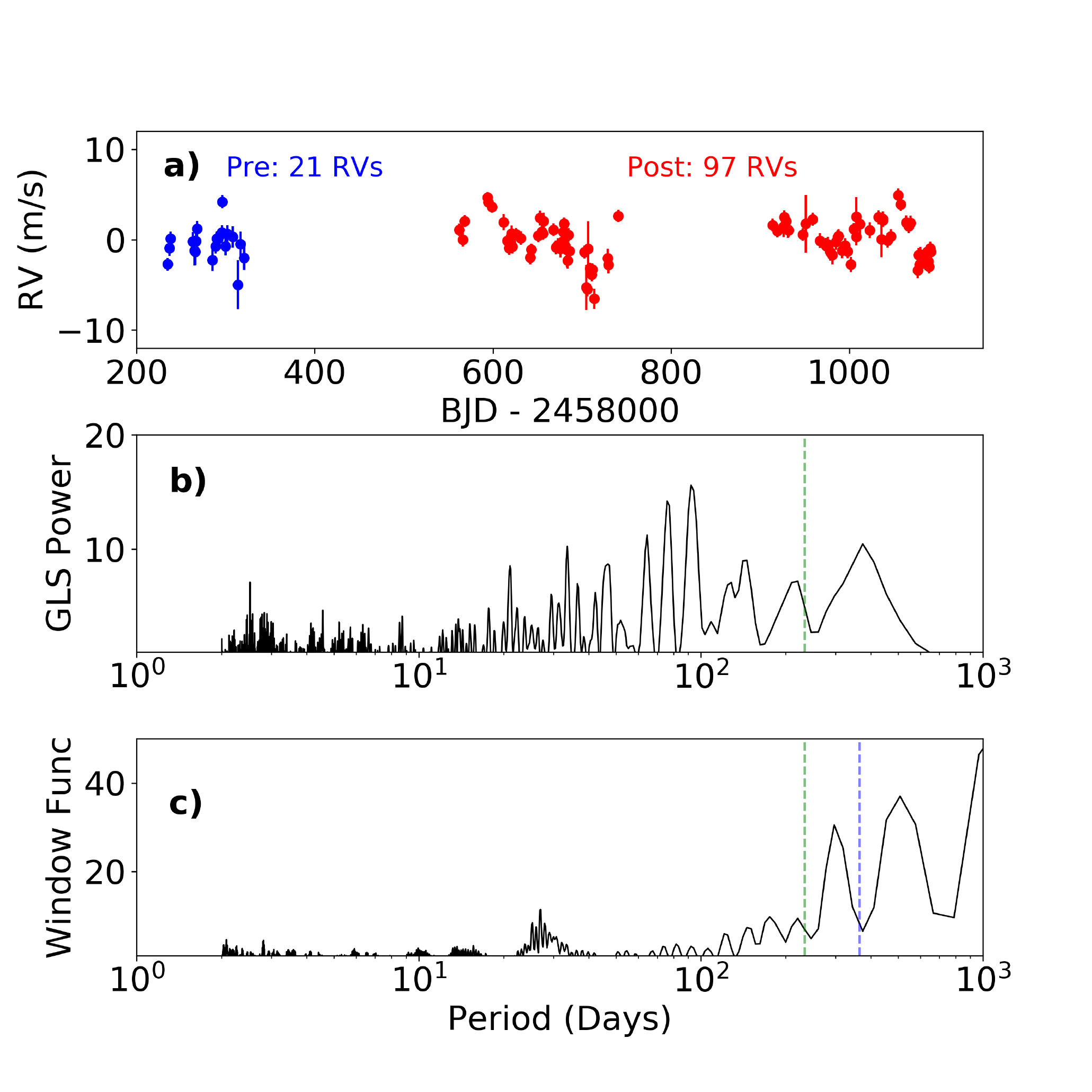}
 \centering\caption{\textbf{Top:} The HPF RV time series split into ``pre" (blue) and ``post" (red) with respect to the thermal break. \textbf{Middle:} The GLS periodogram for the HPF RVs. The green dashed line is at the proposed planet period. \textbf{Bottom:} The accompanying window function periodogram. The blue dashed line denotes 365 days.}
  \label{fig:hpf}
\end{figure}

\par From these early results, we performed a season-by-season, instrument-by-instrument analysis of the discovery data in an effort to identify tighter, more scientifically grounded boundaries for splitting the data. We found that for both the H$\alpha$ and the $S_{HK}$ time series, the rotation period of the star, 145 $\pm$ 15d \citep{ToledoP2019}, is recoverable in the 2011 season by visual inspection when plotting out the data. Through this, we realized that the 233 day signal persisted with significant power only for a 1000 day stretch of time ---from BJD 2455600 to 2456600, 211 observations (see Fig \ref{fig:timeseries} --- during the observing seasons of 2011, 2012, and 2013). For ease of communication, we have subsequently designated this time span as \textit{the 1000d window}. From here on, we begin referring to time with respect to the 1000d window: the before, during, and after epochs. The RV and activity periodograms in Figure \ref{fig:3x3} show the behavior of the 233d and 145d signals in each of our three epochs.

\par When we stitch back together the Before and After RVs of the discovery data set to create a data set with 560 points, albeit with a noticeable gap in the center, we still do not see any significant power at the 233 day period. In fact, the GLS power of the 233 day signal drops from $\sim31$ down to $\sim15$. As GLS power scales exponentially with significance, this loss in power corresponds to a large loss in statistical significance. Additionally, the 233 day signal loses its place as the top peak in favor of a 45 day signal.

\begin{figure}[t]
  \centering\includegraphics[width=.5\textwidth]{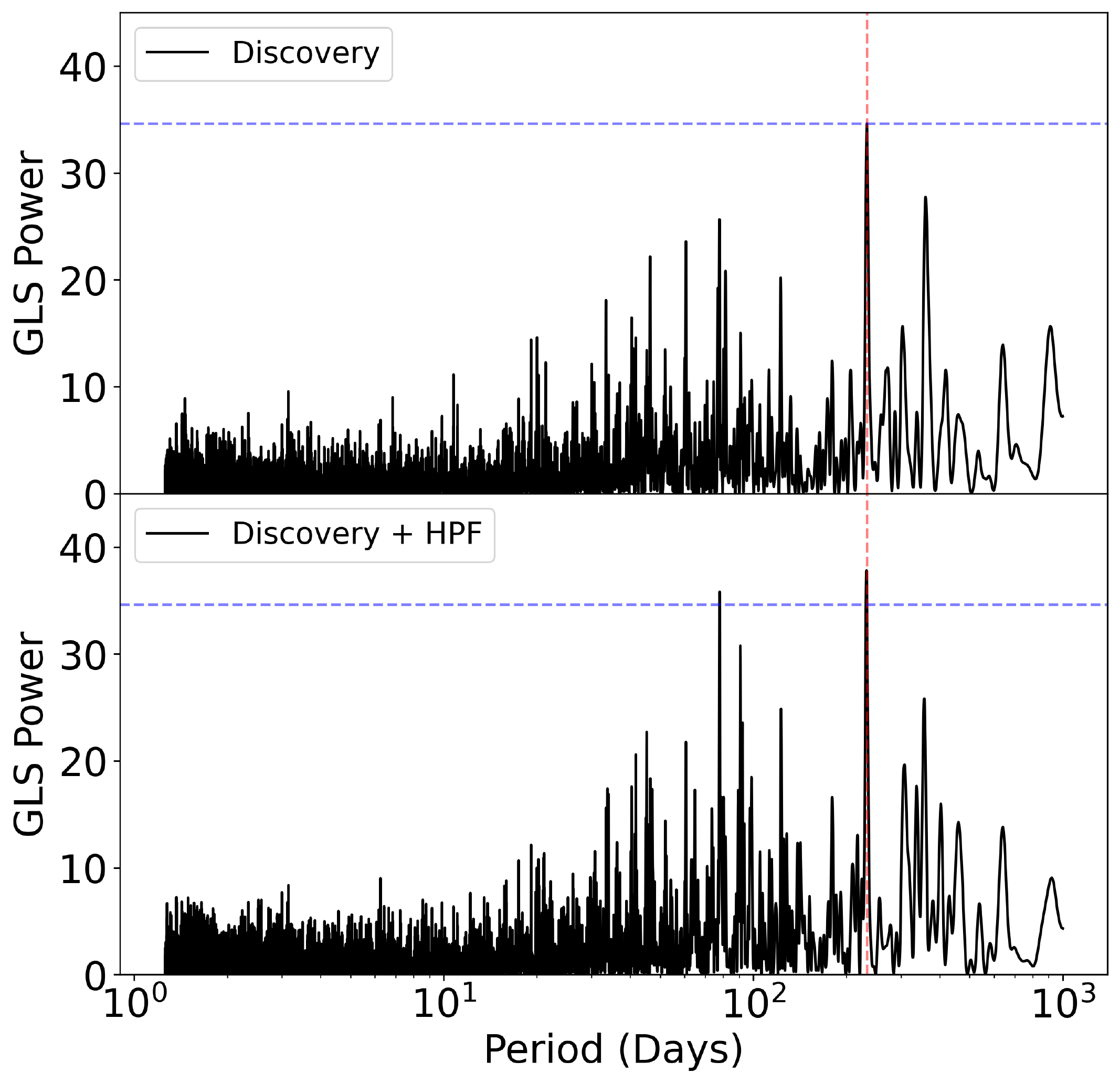}
 \centering\caption{A comparison of the GLS periodograms for two different data sets. The addition of HPF data to the original discovery data set only marginally increases the strength of the peak at 233 days, while the peak at 77 days, the first harmonic of the rotation period, greatly increases. The horizontal blue dashed lines are set at the height of the peak at 233 days in the discovery data set, and are intended to guide the eye.}
  \label{fig:hpfmodelcomp}
\end{figure}

\par With this in mind, we turned to the new HPF data. Figure \ref{fig:hpf} shows the time-series HPF RVs as well as its periodogram and window function. We do not see evidence for a signal at 233 days. This could be due to a combination of the fact that HPF is a NIR spectrograph, which is less susceptible to spot-dominated stellar activity than a visible spectrograph, and/or the star could have been less active in the last few years than it was in 2011-2013. The top peak in the HPF data's GLS periodogram is at 92 days. We believe this peak is part of a comb of aliases of the $\sim$75 day (second peak) first harmonic of the rotation period.

\par With the addition of 118 new data points, we would have expected the power of the signal at the proposed planet's period to increase. Adding the HPF RVs to the discovery data set (note this is not the same as our \textit{updated} data set since it still includes the original reductions of the HARPS and non-split HIRES data sets), we recover the signal at 233 days in a GLS periodogram at nearly the same power as is found when only considering the discovery data alone ($\Delta Power \simeq 3$) while the signal at 77 days has a large increase in power ($\Delta Power \simeq 10$), see Figure \ref{fig:hpfmodelcomp}. As explored in detail in \S \ref{consecremove}, the HPF RVs do not increase the power at 233d as much as would be expected for a genuine exoplanet. When we exclude the 1000d window observations from this combined data set, the 233 day signal again loses significant power and no longer retains its place as the top peak in the periodogram.

\par Lastly, we have chosen to demonstrate the transitory nature of the 233 day signal with the Lomb-Scargle periodogram due to the broader community's common understanding of the algorithm. Regardless of the periodogram we used (Bayes Factor, Marginalized Likelihood \citep{Feng2017}; Compressed Sensing \citep{Hara2017}), when we remove the observations in the 1000d window we see the same result. The strength of the 233 day signal is almost entirely contained within the 211 data points which compose the 1000d window.

\subsection{Sampling Sensitivity}
\label{samplesense}

Such a dramatic reduction in signal significance by the exclusion of certain data prompted us to ask if it was reasonable to remove 27.6\% of the data (the percentage of the RV time series which is inside the 1000d window for the discovery data set) and still expect a signal to persist. In an effort to address this, we created subsets of data where we removed RV points from the time series, and computed the GLS periodogram of the resulting subset.

\subsubsection{Random Removal}
\label{randomremove}

\par First, we experimented with an approach of randomly removing data. We created 2000 different subsets of the discovery data where 211 observations---the same number as are found in the 1000d window---were randomly removed. For the first 1000 subsets, we allowed the randomizer to choose any observations for removal (except for the first and last observation, as we did not want to change the baseline of the time series in any subset). In the second 1000 subsets, we enforced that the randomizer \textit{not} choose any observations that were taken during the 1000d window (as well as first and last observation again). This is an important distinction because it tests how localized the signal is by keeping the data where we believe all of the signal power lies.

\par
\begin{figure}[]
  \centering\includegraphics[width=.52\textwidth]{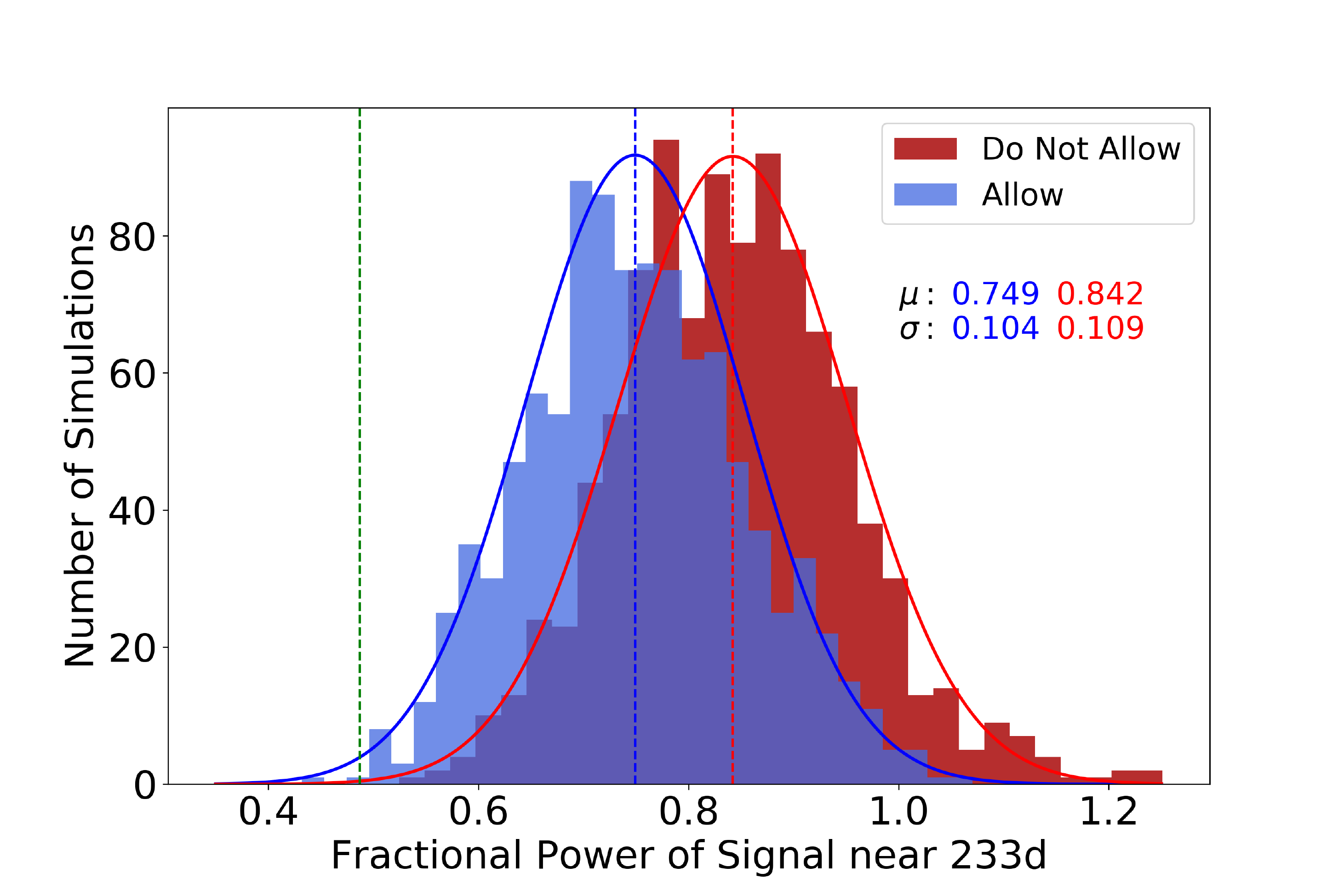}
  \centering\caption{Histograms of the GLS Power from randomly removing 211 observations from the discovery data set in 1000 subsets via two different methods: where we allow for observations within the 1000d window to be part of the random removal (blue) and for when we do not allow those observations to be part of the random removal (red). The green line indicates the subset which corresponds to removing only the 211 observations that make up the 1000d window. Gaussian best fit parameters are shown in their respective colors.}
  \label{fig:random_removal}
\end{figure}

 For each subset, after removing the randomly selected observations, we re-calculated and removed zero-point instrumental offsets and then computed the GLS peridogram. In each subset's periodogram, we tracked how the power of the signal at 233 days was affected by the removal of data. Figure \ref{fig:random_removal} shows the fractional power (normalized by the true power of the signal when all discovery data is included) of a signal peak nearest to 233 days within a $\pm$ 3 day span (in fact, we see the same results when expanding this as far as $\pm$ 20 days) in each of these two rounds of subsets.

\par For the those where we allowed for any observation to be removed, the average loss in power of the 233 day peak is 25\%. By fitting a Gaussian distribution to the ensemble of subsets, we see that when we remove the specific 211 observations which make up the 1000d window, the resulting subset constitutes a 2.5 sigma outlier. For those subsets where we do not allow for removal of any observations from the 1000d window, the average loss of power is less, only 16\%. This result is consistent with our hypothesis that the power of the 233 day signal lies mostly in the 1000d window. A Gaussian fit to this ensemble shows the removal of the specific 1000d window observations is a 3.25 sigma outlier. Furthermore, and equally important, for both methods of removal there are subsets of the discovery data where the significance of the 233 day signal actually \textit{increases}, by up to 20\%. These increases are notable because they indicate there are subsets where the removal of data significantly strengthens the proposed planet signal.

\begin{figure*}[t]
  \centering\includegraphics[width=\textwidth]{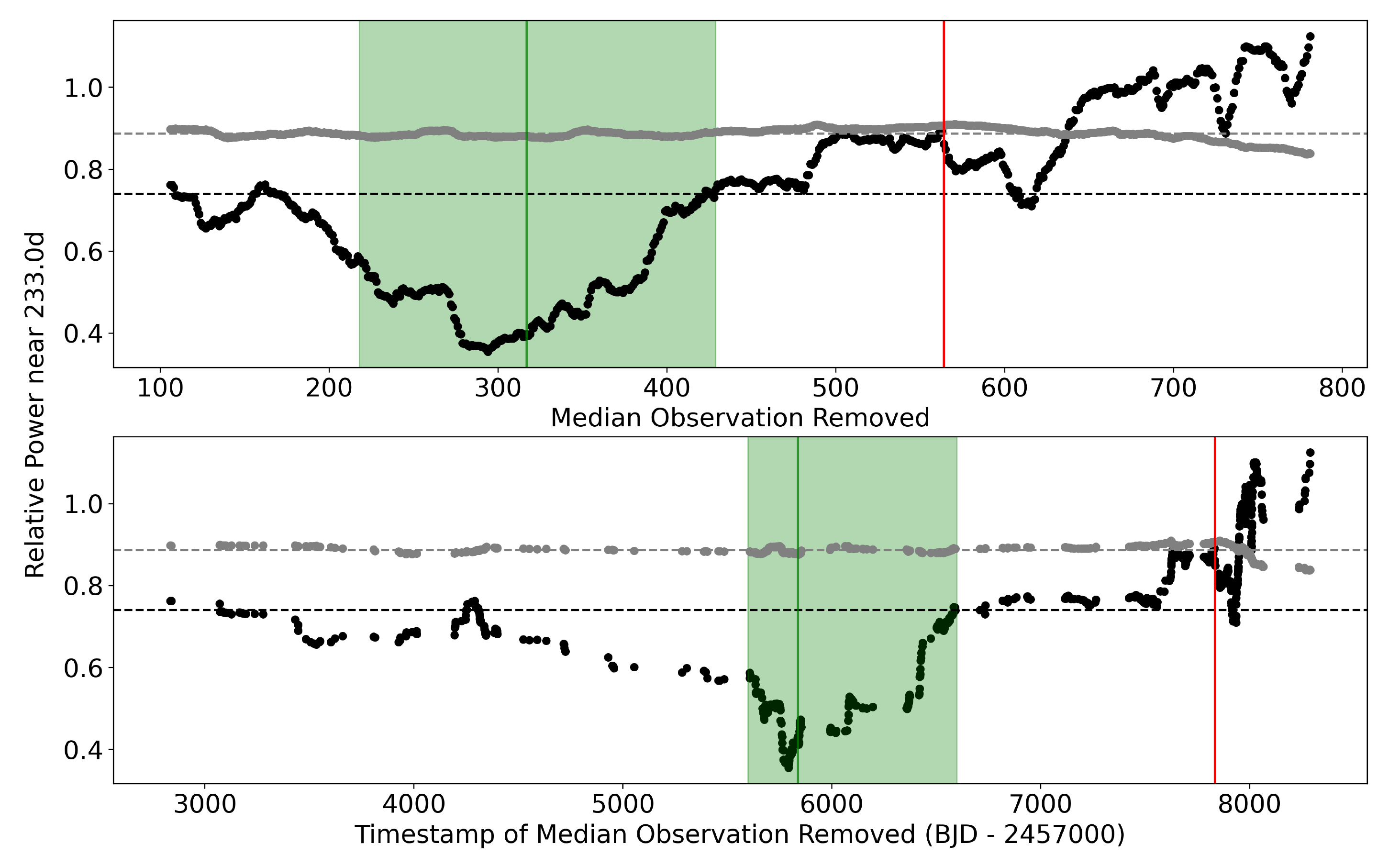}
  \centering\caption{\textbf{Above: } The results from the rolling omission tests in observation number space. Black points are the true updated data set, gray points are an average of 10 synthetic data sets. The green line is the simulation which matches the removal of the 1000d window points exactly, with green shading to indicate the duration of the 1000d window. The red line indicates the simulation where the first observation of the 211 consecutive removed observations corresponds to the first CARMENES observation. \textbf{Below: } The same results but in time space.}
  \label{fig:consecutive_removal}
\end{figure*}

\subsubsection{Consecutive Removal}
\label{consecremove}

\par Next, we performed a similar analysis wherein we removed 211 \textit{consecutive} data points, by a process we call \textit{rolling omission}. First, we fit for instrumental zero point offsets. We created an algorithm to start with an observation, remove the subsequent 211 points, compute the GLS periodogram for the subsequent subset of the data, and track the fractional power of the signal (data subset power at 233d / full data set power at 233d). Then increment to the next starting observation and repeat for the next subsequent 211 observations. Since we did not want to create a subset with a different baseline than the full data set, the first iteration began with the removal of the second observation and the last iteration began with the observation equal to $N - 211 - 1$, where $N$ is the total number of observations in the time series. We performed this test on both the discovery and updated data sets, but the results are more illuminating with the addition of the HPF data in the updated data set, primarily because with more data points, the algorithm is able to run longer before reaching the end.

\par To show the results' deviation from expectation more clearly, we performed the rolling omission test on synthetic data in which we include a planet signature. We created synthetic data sets where for each timestamp and associated error of the real observations, we calculated the corresponding expected velocity for a planet with the same parameters as published in by R18. Then we added a white noise term to each velocity. Noise was drawn from a Gaussian centered at zero with width equal to the error measurement for the observation at that timestamp. We created 10 of these synthetic data sets and ran the rolling omission test on each one, averaging the results.

\par Figure \ref{fig:consecutive_removal} shows the result of the rolling omission test on the updated data set (black points) and the averaged synthetic data sets (gray points) as function of both observation number and time. The vertical green line indicates the iteration of rolling omission where the subset removed matches exactly the removal of the specific 211 observations of the 1000d window. The green shading shows the width of the 1000d window in index and timestamp space. The red vertical line shows the iteration of rolling omission where the first point of the consecutive removal is equal to the first observation in the CARMENES time series. Like in the random removal experiments, this experiment shows that the loss in fractional power when we remove the specific 211 observations that make up the 1000d window is far greater than the average (black dashed line).

\begin{figure*}[t]
  \centering\includegraphics[width=0.9\textwidth]{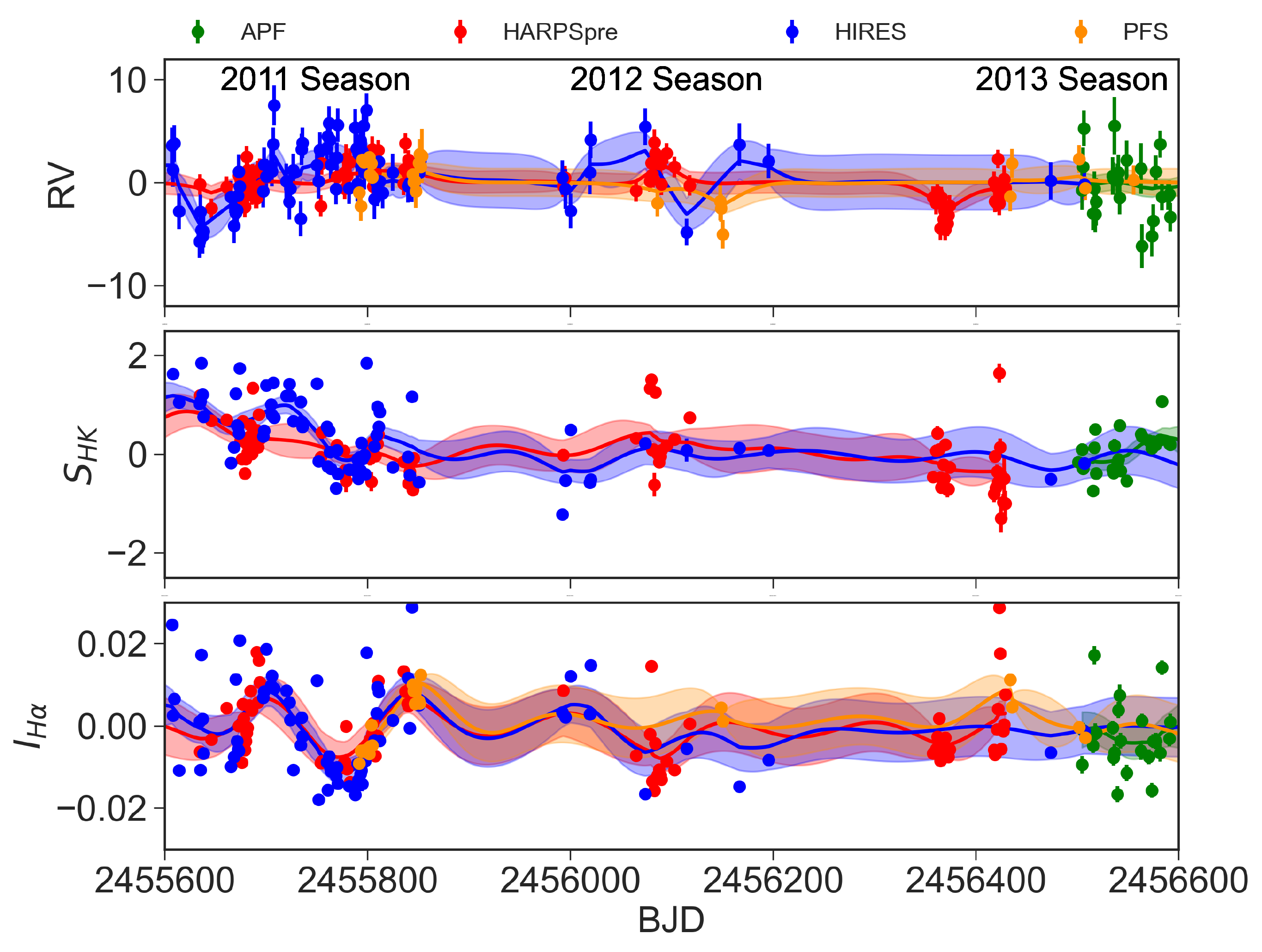}
  \centering\caption{The time series of RV, $S_{HK}$, and $I_{H\alpha}$ in the 1000d window with GP functions overlaid with 1 $\sigma$ uncertainty regions. In the first of these observing seasons, 2011, the rotation period of the star is recoverable by eye in $S_{HK}$ and H$\alpha$.}
  \label{fig:1000d}
\end{figure*}

\par In addition, the tests on synthetic data sets track a nearly horizontal line, showing that for a signal which persists throughout the span of the data the loss in power from removing consecutive chunks of data is nearly agnostic to the window that is removed. When the data is dense in time, as it is toward the end of the data set, the fractional power begins to decrease. However, for the real data set, we see just the opposite behavior: the fractional power \textit{increases} dramatically after the red vertical line, even going above 1.0. This means that as we remove some of the best data, in terms of nightly cadence, RV precision, and redder instruments less susceptible to spot-dominated stellar activity in the case of the HPF and CARMENES data, the signal becomes \textit{stronger} and more significant than if we include all of the data. This is because we are removing data which does not actually contain a signal at 233 days, thus concentrating the subset on the observations which truly contain the signal.

\par Through these tests we became convinced that not only does nearly all of the signal power of the proposed planet lie in the particular 211 RVs that make up the 1000d window, but also that if the signal were persistent in time, our analysis suggests that the most recent observations from CARMENES, HARPS, HARPS-N, and HPF should add power to the 233d signal if it is caused by an exoplanet, when in reality they do not.

\section{Discussion}
\label{discussion}

\par This result is both novel and concerning. It is novel in the sense that there are no examples in the literature of a false positive exoplanet detection created by a stellar rotation alias \textit{longer} than the rotation period. \citet{Boisse2011} showed that confidently disentangling planetary signals from stellar activity induced signals requires a few circumstances, primarily that the period of the planet signal not be near the stellar rotation period or one of its harmonics. Harmonics, by definition being integer multiples of the frequency, are shorter periods. While false positive exoplanet signals have also been identified at rotation period aliases, they too tend to fall towards shorter periods \citep{Robertson667C}, and the same is for true planets where the period is uncertain between two aliases \citep{Dawson2010, Robertson2018}. This result has farther reaching implications for the field of RV exoplanet discovery than just this one system.

\par Equally, this result is concerning. The exoplanet science community will now have to be as vigilant in scrutinizing the longer aliases of the stellar rotation as it has been for shorter aliases and harmonics. Similarly concerning, this result seems to contradict Extended Data Figure 2 in R18 where they show a stacked periodogram in the fashion of \citet{Mortier2017}. The figure shows a monotonically rising confidence in the 233 day signal with increased observation. How can we reconcile this figure with our own Figure \ref{fig:consecutive_removal} where we see the signal is localized to a relatively small time and observation space window? An alias is, in itself, a signal which can increase in significance monotonically in the same manner as a planetary signal. Additionally, the periodogram of the first $X$ observations should achieve similar results to the periodogram of the first $X+1$ observations. A better approach would be to stack periodograms of disjoint time spans of the data, similar to our rolling omission test. Along those lines, our rolling omission test is better suited for tracing the localization of a signal in time, which, when used in concert with a stacked periodogram, can provide a more holistic understanding of the nature and veracity of a signal.

\par We have described the false positive 233d exoplanet signal as an ``alias" of the 142-day rotation period due to its apparent connection to the stellar rotation and the location of the periodicity near the 1-year alias of said rotation: $\big(\frac{1}{145}-\frac{1}{365}\big)^{-1} = 240$. However, this signal cannot be explained as simply as the result of a pure 142-day sinusoid sampled at the irregular cadence and precision of the RV time series. Instead, we believe several factors explain why the rotation signal manifested as observed in RV, H$\alpha$, and $S_{HK}$:
\begin{itemize}

    \item During the 1000d window, the sampling between H$\alpha$ and $S_{HK}$ observations was not identical but also not too different. Yet the resulting periodograms are very different. In the 1000d window, the H$\alpha$ values track a decaying and coherent-across-instruments stellar rotation signal, while the $S_{HK}$ values generally do not, see Figure \ref{fig:1000d}. The $S_{HK}$ shows the rotation period only in the first season, 2011, and in one instrument, HIRES. This might be explained by the low SNR in the continuum of this very red star near the Ca II H\&K doublet \citep{Robertson2016}.

    \item H$\alpha$ and $S_{HK}$ often trace different astrophysical phenomena \citep{Silva2014}.  The activity and periodicity associated with one or the other---or even both---can imprint through to the RVs.

    \item  The decaying aspect of this signal would make aliasing worse. As shown in Figure \ref{fig:1000d}, the rotation signal during the critical 1000-day window is prominent in the first season, and then slowly decays. Irregular sampling of this decaying, non-sinusoidal signal, alongside any unknown true low-amplitude planet(s) might have created the perfect storm resulting in the false positive detection of Barnard's star b.

\end{itemize}

\par It is crucial that we have a strong understanding of Barnard's star if we are to continue using it as a standard star for commissioning and monitoring red-optical and NIR RV instruments. The star is often considered to be \textit{the} RV standard for the mid-to-late M dwarf class. As our spectrometers become capable of breaking through ever smaller instrumental noise floors, we will need a thorough understanding of the astrophysical variability of any standard star. Barnard's star has always been considered a quiet star, and this is still largely true for signals on the order of a few meters per second and up. But we now see that the stellar activity of Barnard's star will play a factor at amplitudes near and below 1 m/s. If we do not properly understand this stellar activity, then we may not be able to use the star as a standard for commissioning new, ever more sensitive instruments in the future.

\par As astronomy enters this new era of highly precise instruments capable of detecting exoplanets with RV amplitudes $<1$ m/s, we are excited by the outlook of finding more low amplitude planets, be they low mass or long period or both. However, long period planets with small amplitudes in particular will need extra care when the proposed period is close to, not only the rotation period of the star or its harmonics, but also its longer seasonal aliases. This will be even more important when the period of a proposed signal is nearly as long as the duration of an observing season.

\par Planetary companions to Barnard's star would be very interesting, not only for their relative proximity to our Sun and therefore enhanced follow-up opportunities, but also because the star belongs to an old stellar population. \citet{Gizis1997} classified the star as an ``Intermediate Population II'' star, which places it somewhere between the halo and the thin disk population of the Milky Way. That said, it would be strange if Barnard's star \textit{did not} host a planet. We know from planet occurrence rate studies like \citet{Dressing2015, Hsu2020} that planets are common at short periods in M Dwarf systems, though clustering of planets in multi systems is also seen in both FGK stars \citep{He2021} and M dwarfs \citep{Ballard2016}. Therefore, we might expect Barnard's star to host a planetary system, although, despite $\sim$850 RVs, we may still not be sensitive enough to detect planets. It is also possible that Barnard's star hosts a nearly face-on planetary system, which would effectively hide any planets from transit or RV searches. Direct imaging with the next generation of large telescopes might help resolve this. R18 states that the separation between the proposed planet and the star would be great enough for Hubble to detect an astrometric signal, which \citet{TalOrastrometry} confirms and goes on to include Gaia DR2 as an instrument capable of making this measurement. If such measurements are undertaken, based on the result we have presented in this paper we would not expect a significant signal corresponding to the planet candidate proposed by R18.

\par With all this in mind, we put effort into searching for a planetary signal in the updated data. We investigated a signal at 45 days, which is seen when removing the 1000d window observations. However, this signal is concernedly close to the second harmonic of the rotation period, it disappears in the residuals of the GP only model, and when treated as a planet it is not favored by the model comparison tests (either with or without a GP).

\section{Conclusion}
\label{conclude}

\par The addition of new data from HPF, as well as a suite of tests on the discovery data set, have shown that the 233 day signal is not planetary in origin, rather a transitory one year alias of the stellar rotation period which took place over a 1000-day time span from 2011 to 2013.

\par In summary:
\begin{itemize}
  \item The 233 day period of the planet candidate proposed by R18 is a one year alias of the rotation period, $\big(\frac{1}{145}+\frac{1}{365}\big)^{-1} = 240$ days. The claimed period of 233 days is well within the $1\sigma$ error bars of the rotation period found by \citet{ToledoP2019}, and consistent with our own analysis of the H$\alpha$ time series.

  \item We created various models of the system, taking into account stellar activity and/or a planet. For the updated data set, we have strong evidence ($\Delta BIC = \sim$160) to reject the planet only model in favor of an activity only model. Then, while we can never rule out one or more planets in the system, our analysis finds that the updated data favors the activity only model over the activity+planet model ($\Delta BIC = \sim$1).

  \item The RV signal at 233 days does not persist through all the tested time windows. Rather, it appears strongly in the middle window, dubbed \textit{the 1000d window} from JD 2455600 to 2456600, comprised of 211 observations, while statistically insignificant in the the other two windows.

  \item The coincidence of the 233 day peak in the RVs and $S_{HK}$ values in the 1000d window while absent otherwise suggests a common stellar activity origin for both signals, aliases of the 145 $\pm 15$ day stellar rotation period.

  \item H$\alpha$ and $S_{HK}$ clearly trace out the stellar rotation period at the beginning of the 1000d window during the 2011 observing season. The H$\alpha$ time series in all windows consistently shows a signal at the stellar rotation period.

  \item Removing observations from a data set will typically weaken any signal, regardless of its astrophysical origin. We created subsets the data by randomly removing 211 observations and find that in nearly all simulations the loss of power resulting from this removal is much less than when we remove the specific 211 observations that make up the 1000d window.

  \item Similarly, we removed 211 consecutive RV observations from the full data set, by a process we call \textit{Rolling Omission} and found that, again, it is the particular 211 RV observations in the 1000d window which have the most impact on the strength of the signal. Removing later chunks of data actually increased the power of the signal.

\end{itemize}

\par The addition of new data can, and should, change our understanding of any system. Barnard's star is listed in the Guaranteed Time Observations (GTO) lists for the ESPRESSO \citep{Pepe2010} and NEID \citep{Schwab2016} exoplanet surveys. The HPF team will also continue to monitor this star. Barnard's star has long fascinated and continually surprised astronomers; we expect this to continue.

\section{Acknowledgments}

\par We thank the anonymous referee for their thorough and helpful comments on this paper. This work was partially supported by funding from the Center for Exoplanets and Habitable Worlds. The Center for Exoplanets and Habitable Worlds is supported by the Pennsylvania State University, the Eberly College of Science, and the Pennsylvania Space Grant Consortium. This work was supported by NASA Headquarters under the NASA Earth and Space Science Fellowship Program through grants 80NSSC18K1114. We acknowledge support from NSF grants AST-1006676, AST-1126413, AST-1310885, AST-1517592, AST-1310875, AST-1910954, AST-1907622, AST-1909506, ATI 2009889, ATI 2009982, and the NASA Astrobiology Institute (NNA09DA76A) in our pursuit of precision radial velocities in the NIR. We acknowledge support from the Heising-Simons Foundation via grant 2017-0494. Computations for this research were performed on the Pennsylvania State University’s Institute for Computational and Data Sciences’ Roar supercomputer. E.B.F. acknowledges the support of the Ambrose Monell Foundation and the Institute for Advanced Study. This work was supported by a grant from the Simons Foundation/SFARI (675601, E.B.F.). The Center for Exoplanets and Habitable Worlds is supported by the Pennsylvania State University and the Eberly College of Science. S.D. acknowledges National Science Foundation AST-1310875 and National Institute of Standards and Technology NIST-on-a-Chip.

These results are based in part on observations obtained with the Habitable-zone Planet Finder Spectrograph on the Hobby-Eberly Telescope. We thank the Resident astronomers and Telescope Operators at the HET for the skillful execution of our observations with HPF, and the HET staff for their dedication to the facility, and enabling these observations. The Hobby-Eberly Telescope is a joint project of the University of Texas at Austin, the Pennsylvania State University, Ludwig-Maximilians-Universität München, and Georg-August Universität Gottingen. The HET is named in honor of its principal benefactors, William P. Hobby and Robert E. Eberly. The HET collaboration acknowledges the support and resources from the Texas Advanced Computing Center.

\facilities{Hobby-Eberly Telescope: Habitable-zone Planet Finder}

Software:
\texttt{Astropy} \citep{astropy2013},
\texttt{barycorrpy} \citep{kanodia2018}
\texttt{celerite} \citep{dfm2017},
\texttt{corner.py} \citep{dfm2016},
\texttt{emcee} \citep{dfm2013},
\texttt{GNU Parallel} \citep{Tange2011a}
\texttt{HxRGproc} \citep{ninan2018},
\texttt{Jupyter} \citep{jupyter2016},
\texttt{matplotlib} \citep{hunter2007},
\texttt{numpy} \citep{vanderwalt2011},
\texttt{pandas} \citep{pandas2010},
\texttt{PyMC3} \citep{pym3reference}
\texttt{radvel} \citep{Fulton2018},
\texttt{SERVAL} \citep{zechmeister2018}

\bibliography{main.bbl}

\appendix

\section{HPF RV extractions}

\par We extract precise RVs from the HPF 1D spectra using the \texttt{SERVAL} template-matching RV-extraction code \citep{Zechmeister2009}, which we have tailored to work for HPF data as discussed in \cite{Metcalf2019} and \cite{Stefansson2020}. To test the robustness of our extracted HPF RVs and accompanying estimated RV uncertainties at these high precision levels, we performed an additional RV extraction test. For this test, we performed an RV extraction where we split each HPF spectral order into four separate segments along the four different readout channels on the HPF H2RG detector. We then compared the standard deviation of the RVs derived from the four different segments to the expected standard deviation from our estimated RV uncertainties. In doing so, we noticed that for some observations the standard deviation across the four different segments was slightly higher than than expected from our RV uncertainty estimates derived from the inherent RV information content in the HPF spectra and RV template. We suspect this increased level of scatter is due to additional sources of systematic noise in the HPF H2RG detector. H2RGs are known to have a number of systematic noise sources that affect precision RV extractions including bias-level fluctuations, persistence effects, and cross-hatch patterns \citep[see e.g.,][for a discussion on some of these effects in the HPF H2RG]{ninan2018}. To account for the additional source of systematic noise we see between the different HPF readout channels, we took the following steps to conservatively increase the estimated RV error bar from the normal SERVAL pipeline.

\begin{itemize}

\item First, as the SERVAL pipeline has been demonstrated to be accurate at the 1m/s level in \cite{Zechmeister2009}, we extract the HPF RVs order by order using the full RV order. This results in an RV value and RV uncertainty estimate per HPF order. We follow \cite{Zechmeister2009} and \cite{Metcalf2019} and perform a weighted average per RV order.

\item Second, we independently reduce the same spectra by splitting each spectrum into disjoint segments along the four HPF readout channels (512 pixel wide). This yields four independent RV estimates and accompanying RV uncertainties for a given observation after performing a weighted average across all of the orders analyzed (independently for each segment of the spectrum).

\item Third, we modeled the four readout channel RVs ($v_i$) as statistically independent draws from a Gaussian distribution with the mean equal to the full order RV ($v_f$). The Gaussian model's width was set to what we would expect from the reduced information in the readout channel level spectra if the expected error in the full order RV ($\sigma_f$) is multiplied by an unknown {\it inflation factor} $k$. i.e,

\begin{equation}
 v_i \sim N(v_f,\sigma_i) ~~~ \mathrm{where,}~~ \sigma_i = C_i \times (k \sigma_f).
\end{equation}

\par The reduction in the information-content ($C_i$)  due to reducing the size of the spectra into four chunks is $\sim$2. This Bayesian model was implemented in the \texttt{pymc3} package \citep{pym3reference} and fitted to each epoch of data. For most of the observations, the inflation factor $k$ was found to be consistent with 1 (implying the scatter in RVs across the four readout channels is consistent with the full order RV and its derived RV uncertainty). For those epochs where the inflation factor $k$ is greater than 1 with a probability of 95\%, we inflated the full order RV uncertainty by the median of the posterior distribution of the inflation factor $k$.

\item Lastly, using the newly derived RVs for all of the epochs we perform a final weighted average of RVs within a given HET track to give a binned RV point and accompanying RV uncertainty per HPF visit.
\end{itemize}

\par
\begin{figure}[h!]
\centering\includegraphics[width=0.7\textwidth]{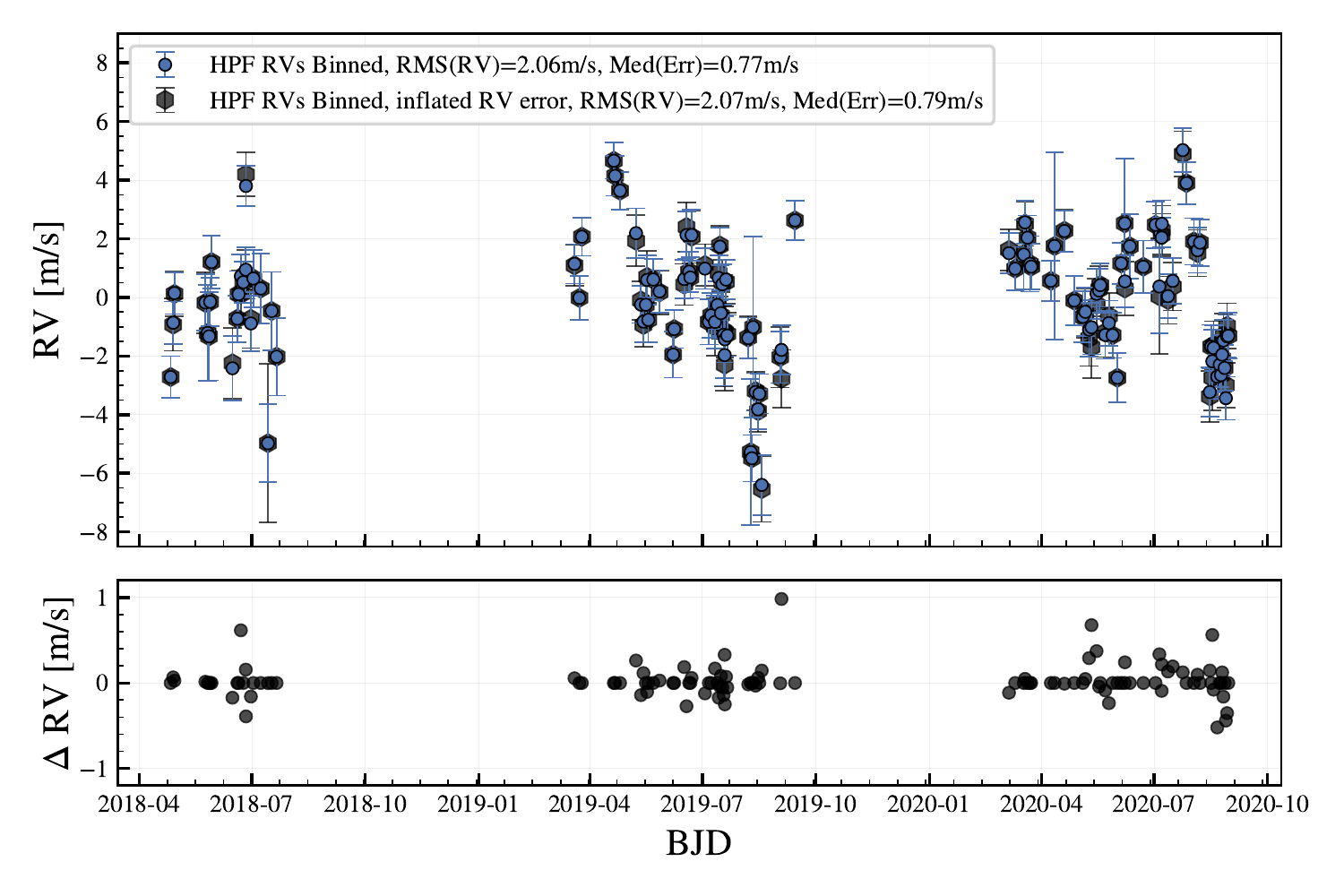}
\centering\caption{HPF RVs of Barnard's Star before and after accounting for additional RV noise between the different HPF readout channels. We use the inflated RV uncertainties for the final RV analysis.}
\label{fig:rvquad}
\end{figure}

 Using this methodology, of the 1016 unbinned observations, we found that 89 observations have sufficient evidence to increase the RV uncertainty by the inflation factor. Figure \ref{fig:rvquad} compares the RVs and the RV uncertainty of the binned HPF RVs before and after applying this correction to the RV uncertainties. In this binned-by-track view, although a few of the observations have their RV uncertainty increased, the median RV uncertainty remains similar before and after applying this correction. As an additional test, we also tried running all of the same analysis presented in this work with the un-inflated RV uncertainties, and achieve the same results.

\section{HPF RVs}

\begin{table}[h!]
\footnotesize
 \centering
 \caption{HPF RV measurements of Barnard's Star.}
 \label{tbl:HPFRVs}
 \begin{tabular}{llll}
    \hline
    \hline
    BJD & RV (m/s) & Err (m/s) & Split \\
    \hline
2458234.87812 & -3.715 & 0.709 & Pre \\
2458236.87809 & -1.93 & 0.893 & Pre \\
2458237.87424 & -0.881 & 0.755 & Pre \\
2458561.98905 & 0.09 & 0.693 & Post \\
2458565.96692 & -1.016 & 0.743 & Post \\
2458567.96441 & 1.071 & 0.648 & Post \\
\hline
\hline
\end{tabular} \\
\footnotesize \textbf{\textsc{NOTES}}\\
\footnotesize The full data set is available online in a machine readable format.
\end{table}

\pagebreak

\section{Additional Tables}
\label{appendix}

\begin{table}[h!]
\footnotesize
 \centering
 \caption{Discovery Data Model Comparison}
 \begin{tabular}{lcc||ccc}
 \hline
 \hline
& \multicolumn{2}{c}{Parameters} &  \multicolumn{3}{c}{$\Delta$BIC}  \\
Model & Free Parameters & BIC & GP only & GP + Planet & Planet only \\
\hline
\textbf{GP only} & \textbf{26} & \textbf{3485.0785} & -- & $\minus$2.040 & 99.814  \\
GP + Planet & 31 & 3487.1186 &  & -- & 97.774 \\
Planet only & 21 & 3584.8938 & & & --\\
    \hline
    \hline
\end{tabular} \\
\footnotesize \textbf{\textsc{NOTES}}\\
Our preferred model, taking into account $\Delta$BIC, is in \textbf{bold}
\label{tbl:disc_compare}
\end{table}

\begin{table}[h!]
\footnotesize
 \centering
 \caption{Updated Data Model Comparison}
 \begin{tabular}{lcc||ccc}
 \hline
 \hline
& \multicolumn{2}{c}{Parameters} &  \multicolumn{3}{c}{$\Delta$BIC}  \\
Model & Free Parameters & BIC & GP + Planet & GP only &Planet only \\
\hline
GP + Planet & 40 & 4021.3900 & -- & 1.354 & 160.970 \\
\textbf{GP only} & \textbf{35} & \textbf{4022.7433} & & -- & 159.616 \\
Planet only & 27 & 4182.3590 & & & -- \\
    \hline
    \hline
\end{tabular} \\
\footnotesize \textbf{\textsc{NOTES}}\\
Our preferred model, taking into account $\Delta$BIC, is in \textbf{bold}
\label{tbl:update_compare}
\end{table}

\begin{table}[h!]
\footnotesize
 \centering
 \caption{Priors and Posteriors for Discovery H$\alpha$ Data Sets}
 \label{tbl:halpha_model}
 \begin{tabular}{lllllll}
    \hline
    \hline
    Model & Parameter & Prior &  Posterior  \\
\hline
GP & $\log L$ & $\mathcal{U}$(0.1, 6) & $1.91_{-0.19}^{+0.19}$ \\ 
 & $\log C$ & $\mathcal{U}$(-6, 6) & $-5.44 _{-0.51}^{+2.60}$ \\
 & P$_{rot}$ & $\mathcal{G}$(145, 15) & $142.96_{-10.49}^{+11.65}$ \\
 & $\log B$ & $\mathcal{U}$(-9, 6) & $-4.28 _{0.08}^{+0.08}$ \\
 Instrument & $\sigma_{APF}$ & $\mathcal{U}$(0.001, 1) & 0.01 $\pm0.001$ \\
 & $\sigma_{CARMENES}$ & $\mathcal{U}$(0.001, 1) & 0.01 $\pm0.001$ \\
 & $\sigma_{HARPSN}$ & $\mathcal{U}$(0.001, 1) & 0.01 $\pm0.001$ & \\
 & $\sigma_{HARPSpre}$ & $\mathcal{U}$(0.001, 1) & 0.01 $\pm0.001$ \\
 & $\sigma_{HARPSpost}$ & $\mathcal{U}$(0.001, 1) & 0.01 $\pm0.001$ \\
 & $\sigma_{HIRES}$ & $\mathcal{U}$(0.001, 1) & 0.01 $\pm0.001$ \\
 & $\sigma_{PFS}$ & $\mathcal{U}$(0.001, 1) & 0.01 $\pm0.001$ \\
 & $\sigma_{UVES}$ & $\mathcal{U}$(0.001, 1) & 0.01 $\pm0.001$ \\
 & $\gamma_{APF}$ & $\mathcal{U}$(-5, 5) & -0.51 $\pm0.001$ \\
 & $\gamma_{CARMENES}$ & $\mathcal{U}$(-5, 5) & -0.52 $\pm0.001$ \\
 & $\gamma_{HARPSN}$& $\mathcal{U}$(-5, 5) & -0.52 $\pm0.001$ \\
 & $\gamma_{HARPSpre}$ & $\mathcal{U}$(-5, 5) & -0.51 $\pm0.001$ \\
 & $\gamma_{HARPSpost}$ & $\mathcal{U}$(-5, 5) & -0.52 $\pm0.001$ \\
 & $\gamma_{HIRES}$ & $\mathcal{U}$(-5, 5) & -0.51 $\pm0.001$ \\
 & $\gamma_{PFS}$ & $\mathcal{U}$(-5, 5) & -0.52 $\pm0.001$ \\
 & $\gamma_{UVES}$ & $\mathcal{U}$(-5, 5) & -0.51 $\pm0.001$ \\
    \hline
    \hline
\end{tabular} \\
\end{table}

\begin{table}[h!]
\footnotesize
 \centering
 \caption{Priors and Posteriors for \textbf{Discovery} RV Data Sets}
 \label{tbl:disc_rv_models}
 \begin{tabular}{lllllll}
    \hline
    \hline
 Model & Parameter & Prior & GP only & Planet only & GP + Planet \\
\hline
 GP & $\log L$ & $\mathcal{G}$(1.91, 0.19) & $1.60_{-0.12}^{+0.15}$ & -- & $1.56_{-0.13}^{+0.18}$ \\
 & $\log C$ & $\mathcal{U}$(-6, 6) & $-5.85_{ -0.13}^{ +0.57}$ & -- & $-5.88_{-0.10}^{+0.33}$ \\
 & P$_{rot}$ & $\mathcal{G}$(142.96, 11.65) & $141.18_{-12.45}^{+12.95}$ & -- & $139.75_{-13.79}^{+14.08}$\\
 & $\log B_{CARMENES}$ & $\mathcal{U}$(-2, 2) & $0.56_{-0.14}^{+0.16}$ & -- & $0.49_{-0.15}^{+0.17}$\\
 & $\log B_{HARPSN}$ & $\mathcal{U}$(-2, 2) & $0.51_{-0.20}^{+0.25} $ & -- & $0.56_{-0.17}^{+0.23}$\\
 & $\log B_{HARPSpre}$ & $\mathcal{U}$(-2, 2) & $0.05_{-0.13}^{0.14} $ & -- & $0.36_{0.19-}^{+0.21}$\\
 & $\log B_{HARPSpost}$ & $\mathcal{U}$(-2, 2) & $0.21_{-0.22}^{+0.30}$ & -- & $0.27_{-0.23}^{+0.28}$ \\
 & $\log B_{HIRES}$ & $\mathcal{U}$(-2, 2) & $0.85_{-0.12}^{+0.12}$ & -- & $0.74_{-0.14}^{+0.15}$ \\
 & $\log B_{PFS}$ & $\mathcal{U}$(-2, 2) & $0.26_{-0.34}^{+0.21}$ & -- & $-0.01_{-0.58}^{+0.23}$\\
 & $\log B_{UVES}$ & $\mathcal{U}$(-2, 2) & $0.70_{-0.13}^{+0.11}$ & -- & $0.67_{-0.13}^{+0.12}$\\
 Instrument & $\sigma_{APF}$ & $\mathcal{U}$(0.5, 10) & $2.95_{-0.45}^{+0.55}$ & $2.86_{-0.44}^{+0.52}$ & $2.88_{-0.48}^{0.60}$ \\
 & $\sigma_{CARMENES}$ & $\mathcal{U}$(0.5, 10) & $1.16_{-0.20}^{+0.19}$ & $1.88_{-0.14}^{+0.16}$ & $1.17_{-0.21}^{+0.20}$\\
 & $\sigma_{HARPSN}$ & $\mathcal{U}$(0.5, 10) & $0.90_{-0.23}^{+0.29}$ & $2.08_{-0.28}^{+0.33}$ & $0.90_{-0.24}^{+0.29}$\\
 & $\sigma_{HARPSpre}$ & $\mathcal{U}$(0.5, 10) & $0.63_{-0.09}^{+0.14}$ & $0.95_{-0.14}^{+0.15}$ & $0.66_{-0.11}^{+0.16}$\\
 & $\sigma_{HARPSpost}$ & $\mathcal{U}$(0.5, 10) & $0.77_{-0.17}^{+0.21}$ & $1.18_{-0.19}^{+0.20}$ & $0.75_{-0.17}^{+0.23}$\\
 & $\sigma_{HIRES}$ & $\mathcal{U}$(0.5, 10) & $1.32_{-0.32}^{+0.30}$ & $2.56_{-0.19}^{+0.20}$ & $1.38_{-0.34}^{+0.31}$\\
 & $\sigma_{PFS}$ & $\mathcal{U}$(0.5, 10) & $0.79_{-0.22}^{+0.39}$ & $1.17_{-0.32}^{+0.34}$ & $0.83_{-0.25}^{+0.41}$\\
 & $\sigma_{UVES}$ & $\mathcal{U}$(0.5, 10) & $0.75_{-0.18}^{+0.28}$ & $2.33_{-0.22}^{+0.25}$ & $0.75_{-0.19}^{+0.29}$\\
 & $\gamma_{APF}$ & $\mathcal{U}$(-10, 10) & $0.47_{-0.52}^{+0.51}$ & $0.21_{-0.47}^{+0.48}$ & $0.19_{-0.47}^{+0.45}$\\
 & $\gamma_{CARMENES}$ & $\mathcal{U}$(-10, 10) & $2.40_{-0.62}^{+0.58}$ & $2.99_{-0.17}^{+0.17}$ & $2.36_{-0.56}^{+0.54}$\\
 & $\gamma_{HARPSN}$& $\mathcal{U}$(-10, 10) & $1.67_{-0.85}^{+0.80}$ & $2.37_{-0.37}^{+0.37}$ & $1.88_{-0.87}^{+0.86}$\\
 & $\gamma_{HARPSpre}$ & $\mathcal{U}$(-10, 10) & $-0.75_{-0.29}^{+0.28}$ & $-0.43_{-0.13}^{+0.13}$ & $-0.61_{-0.23}^{+0.21}$\\
 & $\gamma_{HARPSpost}$ & $\mathcal{U}$(-10, 10) & $2.97_{-0.83}^{+0.82}$ & $3.44_{-0.23}^{+0.23}$ & $3.30_{-0.95}^{+0.88}$\\
 & $\gamma_{HIRES}$ & $\mathcal{U}$(-10, 10) & $0.89_{-0.49}^{+0.46}$ & $-0.65_{-0.23}^{+0.23}$ & $-0.89_{-0.44}^{+0.44}$\\
 & $\gamma_{PFS}$ & $\mathcal{U}$(-10, 10) & $0.38_{-0.42}^{+0.42}$ & $0.36_{-0.29}^{+0.29}$ & $0.25_{-0.33}^{+0.34}$\\
 & $\gamma_{UVES}$ & $\mathcal{U}$(-10, 10) & $1.46_{-0.50}^{+0.49}$ & $1.75_{-0.30}^{+0.30}$ & $1.45_{-0.48}^{+0.49}$\\
 Planet & Period & $\mathcal{G}$(233, 15) & -- & $232.86_{-0.43}^{+0.29}$ & $232.64_{-0.39}^{+0.32}$\\
 & $T_c$ & $\mathcal{G}$(2454937.92, 20) & -- & $2454934.03_{-4.01}^{+4.55}$ & $2454936.11_{-4.33}^{+4.24}$ \\
 & $K_{amp}$ & $\mathcal{U}$(0.3, 10) & -- & $1.16_{-0.12}^{+0.12}$ & $1.16_{-0.23}^{+0.22}$ \\
 & $\sqrt{e}sin\omega$ & $\mathcal{U}$(-1, 1) & -- & $0.53_{-0.13}^{+0.09}$ & $0.59_{-0.22}^{+0.12}$\\
 & $\sqrt{e}cos\omega$ & $\mathcal{U}$(-1, 1) & -- & $-0.16_{-0.18}^{+0.19}$ & $-0.08_{-0.20}^{+0.21}$\\
    \hline
    \hline
\end{tabular} \\
\end{table}

\begin{table}[h!]
\footnotesize
 \centering
 \caption{Priors and Posteriors for \textbf{Updated} RV Data Sets}
 \label{tbl:updated_rv_models}
 \begin{tabular}{lllllll}
    \hline
    \hline
Model & Parameter & Prior & GP only & Planet only & GP + Planet \\
\hline
 GP & $\log L$ & $\mathcal{G}$(1.91, 0.19) & $1.56_{-0.09}^{+0.13}$ & -- & $1.52_{-0.09}^{+0.14}$ \\
 & $\log C$ & $\mathcal{U}$(-6, 6) & $-5.92_{-0.06}^{+0.18}$ & -- & $-5.92_{-0.06}^{+0.16}$ \\
 & P$_{rot}$ & $\mathcal{G}$(142.96, 11.65) & $142.78_{-13.93}^{+13.41}$ & -- & $139.36_{-14.76}^{+14.56}$\\
 & $\log B_{CARMENES}$ & $\mathcal{U}$(-2, 2) & $0.54_{-0.12}^{+0.15}$ & -- & $0.48_{-0.12}^{+0.16}$\\
 & $\log B_{HARPSN}$ & $\mathcal{U}$(-2, 2) & $0.48_{-0.17}^{+0.22}$ & -- & $0.57_{-0.14}^{+0.19}$\\
 & $\log B_{HARPSpre}$ & $\mathcal{U}$(-2, 2) & $0.25_{-0.11}^{+0.13}$ & -- & $-0.11_{-0.13}^{+0.18}$\\
 & $\log B_{HARPSpost}$ & $\mathcal{U}$(-2, 2) & $0.52_{-0.16}^{+0.26}$ & -- & $0.62_{-0.17}^{+0.22}$\\
 & $\log B_{HIRESpre}$ & $\mathcal{U}$(-2, 2) & $0.94_{-0.17}^{+0.19}$ & -- & $0.85_{-0.17}^{+0.19}$ \\
 & $\log B_{HIRESpost}$ & $\mathcal{U}$(-2, 2) & $0.81_{-0.14}^{+0.14}$ & -- & $0.59_{-0.15}^{+0.18}$\\
 & $\log B_{PFS}$ & $\mathcal{U}$(-2, 2) & $0.28_{-0.27}^{+0.16}$ & -- & $-0.01_{-0.31}^{+0.19}$\\
 & $\log B_{UVES}$ & $\mathcal{U}$(-2, 2) & $0.71_{-0.11}^{+0.11}$ & -- & $0.67_{-0.12}^{+0.10}$\\
 & $\log B_{HPFpre}$ & $\mathcal{U}$(-2, 2) & $-1.91_{-0.07}^{+0.19}$ & -- & $-1.93_{-0.06}^{+0.16}$\\
 & $\log B_{HPFpost}$ & $\mathcal{U}$(-2, 2) & $0.64_{-0.11}^{+0.13}$ & -- & $0.64_{-0.11}^{+0.14}$\\
 Instrument & $\sigma_{APF}$ & $\mathcal{U}$(0.5, 10) & $2.97_{-0.50}^{+0.61}$ & $2.88_{-0.46}^{+0.53}$ & $2.89_{-0.50}^{+0.61}$\\
 & $\sigma_{CARMENES}$ & $\mathcal{U}$(0.5, 10) & $1.15_{-0.21}^{+0.20}$ & $1.86_{-0.15}^{+0.16}$ & $1.16_{-0.21}^{+0.20}$\\
 & $\sigma_{HARPSN}$ & $\mathcal{U}$(0.5, 10) & $0.91_{-0.24}^{+0.29}$ & $2.07_{-0.28}^{+0.33}$ & $0.89_{-0.23}^{+0.31}$\\
 & $\sigma_{HARPSpre}$ & $\mathcal{U}$(0.5, 10) & $0.93_{-0.15}^{+0.16}$ & $1.43_{-0.13}^{+0.14}$ & $0.96_{-0.15}^{+0.17}$\\
 & $\sigma_{HARPSpost}$ & $\mathcal{U}$(0.5, 10) & $1.04_{-0.22}^{+0.23}$ & $1.67_{-0.19}^{+0.22}$ & $1.02_{-0.24}^{+0.25}$\\
 & $\sigma_{HIRESpre}$ & $\mathcal{U}$(0.5, 10) & $1.58_{-0.52}^{+0.55}$ & $2.82_{-0.35}^{+0.40}$ & $1.65_{-0.53}^{+0.54}$\\
 & $\sigma_{HIRESpost}$ & $\mathcal{U}$(0.5, 10) & $1.51_{-0.33}^{+0.32}$ & $2.46_{-0.21}^{+0.23}$ & $1.61_{-0.33}^{+0.32}$\\
 & $\sigma_{PFS}$ & $\mathcal{U}$(0.5, 10) & $0.77_{-0.21}^{+0.36}$ & $1.30_{-0.33}^{+0.36}$ & $0.77_{-0.21}^{+0.38}$\\
 & $\sigma_{UVES}$ & $\mathcal{U}$(0.5, 10) & $0.74_{-0.18}^{+0.29}$ & $2.39_{-0.22}^{+0.25}$ & $0.71_{-0.16}^{+0.28}$\\
 & $\sigma_{HPFpre}$ & $\mathcal{U}$(0.5, 10) & $1.38_{-0.31}^{+0.41}$ & $1.31_{-0.31}^{+0.38}$ & $1.27_{-0.32}^{+0.43}$\\
 & $\sigma_{HPFpost}$ & $\mathcal{U}$(0.5, 10) & $0.59_{-0.07}^{+0.14}$ & $1.87_{-0.16}^{+0.19}$ & $0.59_{-0.07}^{+0.13}$\\
 & $\gamma_{APF}$ & $\mathcal{U}$(-10, 10) & $0.48_{-0.50}^{+0.47}$ & $0.23_{-0.47}^{+0.46}$ & $0.28_{-0.42}^{+0.40}$\\
 & $\gamma_{CARMENES}$ & $\mathcal{U}$(-10, 10) & $2.43_{-0.59}^{+0.57}$ & $2.99_{-0.17}^{+0.17}$ & $2.37_{-0.59}^{+0.56}$\\
 & $\gamma_{HARPSN}$& $\mathcal{U}$(-10, 10) & $1.68_{-0.84}^{+0.75}$ & $2.21_{-0.38}^{+0.37}$ & $1.82_{-0.90}^{+0.86}$\\
 & $\gamma_{HARPSpre}$ & $\mathcal{U}$(-10, 10) & $-0.67_{-0.34}^{+0.32}$ & $-0.25_{-0.16}^{+0.16}$ & $-0.50_{-0.26}^{+0.25}$\\
 & $\gamma_{HARPSpost}$ & $\mathcal{U}$(-10, 10) & $-0.86_{-1.01}^{+0.90}$ & $-0.37_{-0.27}^{+0.25}$ & $-0.47_{-0.79}^{+0.92}$\\
 & $\gamma_{HIRESpre}$ & $\mathcal{U}$(-10, 10) & $-1.40_{-0.79}^{+0.79}$ & $-1.55_{-0.43}^{+0.44}$ & $-1.45_{-0.75}^{+0.70}$\\
 & $\gamma_{HIRESpost}$ & $\mathcal{U}$(-10, 10) & $-0.34_{-0.47}^{+0.45}$ & $-0.14_{-0.23}^{+0.23}$ & $-0.32_{-0.40}^{+0.35}$\\
 & $\gamma_{PFS}$ & $\mathcal{U}$(-10, 10) & $0.34_{-0.39}^{+0.38}$ & $-0.37_{-0.30}^{+0.30}$ & $0.26_{-0.31}^{+0.33}$\\
 & $\gamma_{UVES}$ & $\mathcal{U}$(-10, 10) & $1.46_{-0.49}^{+0.50}$ & $1.78_{-0.30}^{+0.30}$ & $1.44_{-0.50}^{+0.48}$\\
 & $\gamma_{HPFpre}$ & $\mathcal{U}$(-10, 10) & $-1.20_{-0.42}^{+0.39}$ & $-0.90_{-0.40}^{+0.39}$ & $-0.99_{-0.43}^{+0.42}$\\
 & $\gamma_{HPFpost}$ & $\mathcal{U}$(-10, 10) & $0.51_{-0.59}^{+0.59}$ & $-1.04_{-0.22}^{+0.22}$ & $-0.58_{-0.62}^{+0.54}$\\
 Planet & Period & $\mathcal{G}$(233, 15) & -- & $231.93_{-0.30}^{+0.49}$ & $232.36_{-0.22}^{+0.17}$\\
 & $T_c$ & $\mathcal{G}$(2454937.92, 20) & -- & $2454920.91_{-7.01}^{+6.29}$ & $2454935.87_{-2.96}^{+2.30}$\\
 & $K_{amp}$ & $\mathcal{U}$(0.3, 10) & -- & $1.19_{-0.13}^{+0.13}$ & $1.28_{-0.28}^{+0.28}$ \\
 & $\sqrt{e}sin\omega$ & $\mathcal{U}$(-1, 1) & -- & $0.10_{-0.23}^{+0.20}$ & $0.60_{-0.21}^{+0.12}$\\
 & $\sqrt{e}cos\omega$ & $\mathcal{U}$(-1, 1) & -- & $0.28_{-0.25}^{+0.16}$ & $-0.24_{-0.16}^{+0.21}$\\
    \hline
    \hline
\end{tabular} \\
\end{table}

\end{document}